%% file: ms.tex
\documentclass[preprint]{aastex}
\usepackage{amsmath}
\usepackage[letterpaper,margin=1.05in]{geometry}
\usepackage{bm}
\usepackage{ulem} 

\newcommand{\rtp}[1]{\ensuremath{^{#1}}}
\newcommand{\abs}[1]{\ensuremath{\mid #1 \mid}}
\newcommand{\apx}{\ensuremath{\sim}}
\newcommand{\pmt}{\ensuremath{\pm}}

\newcommand{\degree}{\ensuremath{^\circ}}

\newcommand{\mas}{\ensuremath{\mbox{mas}}}

\newcommand{\asyr}   {\ensuremath{\mbox{as} \; \mbox{yr}^{-1}}}

\newcommand{\masyr}   {\ensuremath{\mbox{mas} \; \mbox{yr}^{-1}}}
\newcommand{\masyrsq} {\ensuremath{\mbox{mas} \; \mbox{yr}^{-2}}}
\newcommand{\kms}     {\ensuremath{\mbox{km\,s}^{-1}}}

\newcommand{\kmsyrsq} {\ensuremath{\mbox{km\,s}^{-1} \; \mbox{yr}^{-2}}}

\newcommand{\Msun}    {\ensuremath{M_{\odot}}}

\newcommand{\uth}{\ensuremath{^{th} \;}}
\newcommand{\cf}{{cf.}}
\newcommand{\ie}{i.e.,\ }
\newcommand{\eg}{e.g.,\ }
\newcommand{\etal}{et al.}
\newcommand{\gl}{{\ell}}                 
\newcommand{\gb}{b}
\newcommand{\muvec}{\vec{\bm\mu}}
\newcommand{\as} {\ensuremath{''       \;}}
\newcommand{\am} {\ensuremath{'        \;}}
\newcommand{\ad} {\ensuremath{^{\rm o} \;}}
\newcommand{\asb}{\ensuremath{''       \!\!.}}
\newcommand{\amb}{\ensuremath{'        \!\!.}}
\newcommand{\adb}{\ensuremath{^{\rm o} \!\!.}}

\newcommand{\snROv}[1]{\renewcommand{\baselinestretch}{#1}\begin{normalsize}}
\newcommand{\enROv}{\end{normalsize}\renewcommand{\baselinestretch}{1.0}}

\newcommand{\ssROv}[1]{\renewcommand{\baselinestretch}{#1}\begin{small}}
\newcommand{\esROv}{\end{small}\renewcommand{\baselinestretch}{1.0}}

\newcommand{\sfROv}[1]{\renewcommand{\baselinestretch}{#1}\begin{footnotesize}}
\newcommand{\efROv}{\end{footnotesize}\renewcommand{\baselinestretch}{1.0}}

\slugcomment{Accepted to ApJ Supplements}

\shorttitle{Very Wide Binaries and other Comoving Stars}
\shortauthors{Shaya and Olling}

\begin{document}

\pagestyle{plain}
\setcounter{page}{0}

\title{Very Wide Binaries and Other Comoving Stellar Companions: A Bayesian
  Analysis of the Hipparcos Catalogue}

\author{ Ed J. Shaya\rtp{1} \& Rob P. Olling\rtp{1}}

\affil{
\rtp{1}University of Maryland at College Park, Department of Astronomy, 
College Park, MD 20742; eshaya@umd.edu
}

%
%

%
\begin{abstract}

We develop Bayesian statistical methods for discovering and assigning
probabilities to  non-random (\eg physical) stellar companions.  
These companions are either presently bound or were previously bound.  
The probabilities depend on similarities in corrected proper motion parallel 
and perpendicular to the brighter component's motion, parallax, and the local 
phase-space density of field stars. Control experiments are conducted to
understand the behavior of false positives.  The technique is applied to the
Hipparcos Catalogue within 100 pc.  This is the first all-sky survey to locate
escaped companions still drifting along with each other.
In the $<100$ pc distance range, $\sim$220 high
probability companions with separations between 0.01 -- 1 pc are found. 
The first evidence for a population ($\sim$300) of companions separated by 
1 -- 8 pc is found. 
We find these previously unnoticed naked-eye companions (both with $V<6$\uth mag): 
Capella \& 50 Per, $\delta$ Vel \& HIP~43797, 
Alioth ($\epsilon$~UMa), Megrez ($\delta$~UMa) \& Alcor, 
$\gamma$ \& $\tau$~Cen, $\phi$~Eri \& $\eta$~Hor, 62 \& 63~Cnc, 
$\gamma$ \& $\tau$~Per, $\zeta$  \& $\delta$~Hya, 
$\beta^{01}$, $\beta^{02}$ \& $\beta^{03}$~Tuc, N~Vel \& HIP~47479, 
HIP~98174 \& HIP~97646, 44 \& 58 Oph, s Eri \& HIP~14913,
and $\pi$ \& $\rho$ Cep.  
High probability fainter companions ($>$ 6\uth mag) of
primaries with $V<4$ are found for: 
Fomalhaut ($\alpha$~PsA), $\gamma$~UMa, $\alpha$~Lib, Alvahet ($\iota$~Cephi), 
$\delta$~Ara, $\beta$~Ser, $\iota$~Peg, $\beta$ Pic, $\kappa$~Phe  and
$\gamma$~Tuc.

\end{abstract}
\setcounter{page}{1}

\section{Introduction}
\label{sec:Introduction}
The observed binarity and multiplicity rates of stars are significant
clues to star formation processes and galactic dynamics.  For example,
the mass-ratio distribution among pre-main-sequence binaries\footnote{In
 this paper we loosely use the term ``binary''
  to indicate any system that contains more than one star, unless
  otherwise stated.}
indicates that fragmentation rather than common accretion is the dominant
formation process \citep{Goodwin_ea_2007}.  Very wide binaries present a
special opportunity because their separations are larger than the size
of typical prestellar cores and thus are important for understanding
the arrangement of separate stellar disks in low-densities
star-formation sites \citep{Parker_ea_2009}.

After formation, the evolution of binaries is determined by dynamical processes.  
In high to moderate density
environments, most pairs with separations of a few hundred to a few
thousand AU are broken up within a few million years \citep{Parker_ea_2009}. 
Outside of these high density regions, Galactic tides and weak interactions
with passing stars peel off stars with separations
of a few times 10,000 AU on a time scale of about 10 Gyr
(\citep{Heggie_1975, Weinberg_ea_1987}).

Until quite recently, stars were commonly assumed to quickly leave
the scene once they become unbound.  However, recent simulations
find that escaping stars drift apart with relative velocity 
$\lesssim$ 1 \kms\ and remain within a few 100 pc of the primary for
billions of years \citep{Jiang_tremaine_2010}. 
In these simulations the large scale potential is
dominated by Galactic tides, while local perturbations to the
large-scale potential are dominated by stars.
Their model does not include molecular clouds, spiral arms, 
or dark-matter subhalos.  
The simulations indicate that the binarity rate decreases with separation 
out until several tidal radii, at which point the rate actually 
increases and peaks at 100 -- 200 pc.  
In addition, since the Galactic gravity field dominates the trajectories of escaped stars,
they travel along with their ex-primaries, trailing or leading
at roughly constant Galactic radii (like a tidal stream, but of only two or three stars).

On the other hand, if, locally, there are many dark matter
subhalos, companions would be more quickly torn away 
and evidence of previous binaries would be lost.
Thus, observational determination of the frequency and ages of escaped companions in
similar orbits should tell us much about the small scale structure of the Galactic 
gravity  field,
the Oort A- and B-constants, and place stringent constraints on dark matter subhalos. 
 
Existing double star catalogs, such as the WDS \citep{WDS}, are
mostly populated with systems selected using fairly simple criteria
such as proximity in the plane of the sky or common proper motions
(CPM) of high proper motion stars.  Very often, pairs await
confirmation of orbital motions before being accepted as a physical
pair which, obviously, selects against wide binaries.  
In a work based on double stars extracted from a large number of catalogs,
one of us found that the {\it apparent} binarity rate changes dramatically
both with distance from the Sun, and with apparent magnitude
\citep{O2005KB}. 
In that work many catalogs\footnote{
    \label{foot:Used_Catalogs}
The catalogs used were: the Hipparcos Catalogue (HIP);
Tycho-2 Catalogue (TY2) \citep{TYCH2};   Tycho Double Star
Catalogue \citep{TDS}; Geneva-Copenhagen Solar
Neighborhood Radial Velocity Survey \citep{GCSN};
9\rtp{th} Catalog of Spectroscopic Binaries \citep{SB09}; 
4\rtp{th} Catalog of Interferometric Measurements of Binary
Stars \citep{WI4}; Washington Double Star Catalog
\citep{WDS}; and Extra-Solar Planets \citep{exop}. In
addition, the updated parallax information from the new
Reduction of the Hipparcos Catalogue [HIP2, \citet{HIP2}] is used rather
than the values listed in HIP.}
were combined, and about 90\% of HIP stars within 10 pc were
found to be either part of a multiple system and/or an exoplanet
host.  In contrast, only \apx14\% of HIP stars are listed as
multiple. Furthermore, only \apx2\% of HIP stars have
other HIP stars as possible companions.
Indeed, it appears that the completeness of catalogs of field
binaries are very seriously affected by selection
effects [\citet{H1990, O2005KB, Kouw_2006_PhDT, Eggleton_2008,
Kouw_2009_Pairings}]. 
This is especially true for wide binaries for which confusion due to 
field stars is severe.

\subsection{Why Very Wide Binaries?}
 \label{sec:Why_Wide_Binaries}

\citet{Parker_ea_2009} indicate that ``hard
binaries'' with semi-major axis ($a$) $\la 50$ AU are
almost never affected by dynamical processes in either the field or 
inside clusters, while ``intermediate binaries'' ($ 50 \la a \la
1,000$ AU) can be highly affected by dynamical processes, especially if
they are formed in dense star clusters. Unevolved ``wide binaries''
with $a > 1,000$ AU can only have formed in
low density starforming regions with densities less than a few
stars per pc\rtp{3}, the so-called ``isolated star formation'' mode
\citep{Goodwin_2010}.
However, of order 15\% of G-type dwarfs are found in wide binary 
systems, with $a \geq 10^4$ AU, which even exceeds the size of 
isolated star formation regions.
It is thought that systems of this size can only 
form during the dissolution phase of low density clusters 
\citep{Kouw_WideBinForm_2010}.

From a Galactic dynamics perspective, one can expect binary stars with 
separation up to about the tidal or Jacobi radius, ($r_J$), while their relative velocities should be 
about the Jacobi velocity ($v_J$):
\begin{eqnarray}
r_J & = & \left(  
   \frac{G\, (M_1 + M_2)}
        {4\, \Omega\, A} \right)^{1/3} \, \, \sim 1.7\, {\rm pc} \, \, ,
   \label{eq:r_jacobi}\\
   v_J & \approx & 0.05 \, \left(  
   \frac{M_1 + M_2}
        {2} \right)^{1/3} \, \, \kms \, \, ,
   \label{eq:v_jacobi}
\end{eqnarray}
\citep{Jiang_tremaine_2010}, where $G$, is the gravitational constant,
$\Omega$ the angular velocity of the Galaxy at the Solar circle, $A$
is Oort's $A$ constant, and $M_1$ and $M_2$ are the masses of the
components.  
The value of 1.7 pc is valid for the canonical values for
the Galactic constants and individual masses of 1 $\Msun$. Note that the
tidal radius depends weakly, only as the cube root, of total mass.
Thus, if it is the case 
that systems that become unbound remain ``close companions'' for a very
long time, then  the region where bound or almost bound systems can be
found would extend much farther than has been previously suggested
[0.1 -- 0.2 pc; \eg \citet{Heggie_1975, BS_1981, RK_1982,
Weinberg_ea_1987, Quinn_ea_2009}].  
This all suggests that separations from 10,000 AU to several parsecs is in
need of much further study.

\subsection{Some Previous Searches for Very Wide Binaries}
 \label{sec:Searching_for_Very_Wide_Binaries}

The search for companions of high proper motion pairs in 
astrometric catalogs is ongoing; {\eg \cite{L2005} use the
USNO-B1 catalog \citep{USNO-B1}, \cite{GK2004} use the USNO-B1 and
2MASS \citep{2MASS}, \citet{MZH_2008} use the
NOMAD\footnote{
\label{foot:NOMAD}
The NOMAD catalog is a compiled catalog containing positions,
proper motions, some optical colors, {\it and} NIR colors from
2MASS, if available. 
The astrometric data listed comes from the HIP, TY2, UCAC2 and USNO-B catalogs. }
catalog, while \citet{LS2005} and \citet{Raghavan_PhD_2009} use their
own surveys}.  
However, the best astrometric catalogs, HIP, TY2,
UCAC2, and now UCAC3 \citep{UCAC3}, have somewhat escaped the
attention of searches for CPM pairs.  
We note that Caballero (2009, 2010) has embarked on a program to identify
very wide binary systems. 
His earlier work concentrates on {\it common} proper
motion systems with separations
over 16\amb65 in the WDS, while the latter work focuses on
the $\alpha$ Lib + KU Lib system with a separation of 2\adb6 (1.05 pc). 
In these works, the focus has been on {\it common} proper motions. 
However, to search for the widest bound systems and for recently escaped
companions, one must look at separations $>$ 1 pc, 
which corresponds to several degrees of separations for stars within 100 pc.  
But, projection effects cause companions with similar space velocities to 
have dissimilar proper motion and radial velocities, hence at very
wide separations, {\it common} 
proper motion studies will miss true companions and may even lead to misidentifications.
In \S\S\ref{sec:Geometric_ProperMotion_Differences}, we describe how to take these geometric effects into account.
  
Finding companions of nearby stars by rigorous statistical
analyses is a fast way to discover additional nearby stars.  
It may also be a means of discovering more nearby brown dwarfs and white dwarfs,
provided many faint candidate stars are included, as in the larger Tycho-2 (TY2)
or UCAC3 catalogs. 
Finally, the discovery of a substantial number of
late-type stars that are paired to higher mass stars can help
significantly in establishing the metallicity and temperature
scales for these low-mass systems (presumably, both components have
the same [Fe/H]).

Therefore,  it would be fruitful to attempt to
construct a statistically robust catalog of astrometric companions
(both bound systems and escaped binary components) with well
defined selection criteria by datamining modern astrometric catalogs.  
The astrometric catalogs such as HIP, TY2, 
UCAC3 and NOMAD provide order of magnitude better proper
motions than previous catalogs  [but see the cautionary notes by
\citet{MZH_2008} on possible systematic errors for faint NOMAD
sources].  
In this paper, a methodology based on
Bayesian statistics is discussed and applied to the stars in just the HIP
that are within 100 pc.  
In a future work, we will present
results of applying these techniques to stars in the  TY2
that may be companions to stars in HIP.

Throughout this paper, we use ``$d$'' for distance (in pc),
 ``$r$'' for 3-d radial separation,
``$\pi$'' for parallax, ``$\mu$'' for proper motion, 
``$\theta$'' for angular separation, and ``$a$'' for semi-major
axis, unless otherwise stated. The sub- and superscripts ``$p$,'' ``$c$,''
 and ``$f$'' are used to refer to properties of
primaries, companion candidates and stars in ``the field,''
respectively.

\section{Multiplicity: Recent Developments}
    \label{sec:Multiplicity_Recent_Developments}

In their study, \citet{DM1991} (hereafter DM91) use their radial velocity
data in combination with existing astrometric binaries to determine the
distribution of periods of main-sequence G stars within 22 pc.
They find that the parent distribution function (PDF) of periods is
approximately Gaussian in the logarithm of the period,

\begin{eqnarray}
\mathrm{PDF}({\cal P}) &\propto& 
       e^{-\onehalf \left( \frac{\log_{10}{{\cal P}}\, - \, 4.8} {2.3} 
                    \right)^2 }
       \,\,\, ,
       \label{eq:PDF_P_DM1991}\\[-1.5em]\nonumber
\end{eqnarray}
where ${\cal P}$ is the period in days. 
The peak is at ${\cal P}=173$ years ($\sim 35$ AU), while
the 1-$\sigma$ boundaries are 316 {\it days} ($\sim 1$ AU)
 and 34,000 {\it years} ($\sim 1,212$ AU).
From their data, DM91 estimate a binary fraction of $\sim$67\%.  
DM91 also find that their observations are consistent with the assumption
that secondaries are drawn randomly from the initial mass function
(IMF) below the primary; therefore, companions are usually considerably fainter
than the primary. 
There is debate in the
literature over the exact shape of the PDF: DM91's Gaussian
shape was first proposed by \citet{K1942} versus \"{O}pik
classical power-law distribution \citep{O1924}.  Because the integral
of the \"{O}pik-power-law distribution diverges it {\it must} break
down.  This is indeed observed at small and large separation (\eg
\citet{GML2003, CG2004, LB2006}, and references therein).

\citet{Raghavan_PhD_2009}, in his dissertation, presents an impressive
body of work on a sample of stars   that
  significantly extends the DM91 sample.  He scrutinized 454 Sun-like
stars within 25 pc by pulling together up-to-date radial velocity
surveys and the best Hipparcos astrometry, while he also performed a
large survey with the CHARA interferometer for close-in binaries.
Following in the footsteps of recent wide-binary searches, he
``blinked'' between early- and late-epoch sky-survey images, out to
radii of about 10\am, or 10 kAU.  He concludes that (55 \pmt\ 3)\% of
stellar systems are single stars.
For the 25pc sample, we search for companions with separations 
up to 120 times larger than those
blinked by \citet{Raghavan_PhD_2009}. 

\section{Methodology}
\label{sec:Methodology}

Although astronomers have been searching for and finding physically
associated pairs of stars since the time of Galileo, there have not
been thorough studies to explicitly assign probabilities of
association.  In this pilot project, we search for common
proper-motion stellar multiples out to very large separations, as far
as is practical, and set probabilities for these to be more than
merely coincidental.  We do {\it not} limit ourselves necessarily to
high proper-motion pairs as has been done in the past \citep{GC2004,
  LB2006}: although stars with field stellar density
exceeding some threshold in a 5 dimensional box given by distance,
plane of the sky positions and proper motions are dropped.  A region about 
the provisional primary star that is the same extent in sky coordinates as
the field selection region but much smaller in proper motion is used
to provide high quality candidate companions.  Where the field density
is small, any star appearing in this small region has high probability
of being physically associated.  As the field density increases, false
positive detections grow and eventually swamp true companions.  In
the range between these, it should be possible, using control
experiments, to at least provide upper limits to the number of real
companions along with a set of candidates each with moderately low
probabilities.  These candidates can be followed up with radial
velocity measurements to further assess their true nature.

The results of the Hipparcos space-based mission provide high
precision proper motions and parallaxes using only its 3.5 year
baseline.  For many cases, though, the best proper motions are
obtained from catalogs that combine data from several astrometric
catalogs, spread over up to 100 years.  
In our current study, we use the proper motions from TY2 
when available, and from HIP2 otherwise. 
Although the HIP2 errors are often smaller than the TY2 errors, it is 
important to use proper motions over a longer baseline than that of HIP2 to 
ensure that the barycentric motion is used.  
A tight secondary at a few AU could induce proper motions of the primary at 
several times the HIP2 error.
In the DM91 distribution, since about one-half of all binaries have periods
$< 173$ yr ($a \la$35 AU) then, within 50 pc, the orbital motions are
several to tens of \masyr, significantly larger than the proper motion
errors.
Thus, the 
longer time baseline of TY2 makes its proper motions less susceptible than 
 HIP2 to orbital  motions induced by small separation companions.
One magnitude below their respective completeness limit, HIP2 has
errors in proper motion of $\epsilon_{\mu} \sim 0.8$ \masyr\ at
V\apx8.5, while TY2  has $\epsilon_{\mu} \sim 3.5$ \masyr\ at
V=11.5.

As each cataloged star is considered as a primary, all stars
within a radius of $\theta_{outer}$ and more than $\theta_{inner}$
(Table~\ref{tbl:params}), 
within a distance range $\abs{d-d_p}\, <\, \Delta d_{max}$, 
and with $ \Delta \mu\, =\, \abs{\vec{\bm\mu}_c - \vec{\bm\mu}_p}\, <\, \Delta \mu_{outer}$ are selected.  
Of those stars,
the number outside of $\Delta \mu_{inner}$ define the {\it local}
star-density $\rho_f$.  Those stars within a specific proper motion
difference, $\Delta \mu_{lim}$, become candidates for companions.
This value is chosen by examining the simulation and finding a value
that lets in roughly 90\% of the simulated binaries.  One can 
change this parameter to either somewhat reduce the false positive
rate or to allow in lower probability candidate  companions.

\subsection{Bayesian Statistics}
    \label{sec:Bayesian}

Our procedure for determining the probability that two stars are
physical companions relies on observed proper motion differences,
$\Delta \vec{\mu} = \vec{\mathbf{\mu}}_p-\vec{\mathbf{\mu}}_c$, 
angular separation $\theta$ and positional differences of
stars within a chosen range of brightness.  
We are not necessarily looking for bound systems; rather we seek systems that are
unlikely to be the results of random distributions.  
Radial velocity differences are not a metric in these statistics because presently
the fraction of stars with known radial velocities is small except for
very nearby bright stars and because a star's radial velocity can
be perturbed by a close companion.  
It is quite typical for
spectroscopic binaries to have offsets in measured radial
velocities by 20 \kms\ from the true barycentric velocities.  With
time the barycentric velocities can be determined by averaging,
but often this is not yet adequately done.  
However, for cases in which radial velocities are known and where the
barycentric motion is determined, radial velocities can be used to
assess the statistical goodness of the technique of probability
assignment.
   
All stars within a specified range in the observables are
considered to be candidate companions or simply ``candidates",
provided that they are fainter than the provisional primary star.
This jargon is chosen for simplicity of bookkeeping, even though, of course,
the brightest star in a system is not
always the most massive component.

Bayesian statistics can provide an estimate of the probability of
association for each candidate.  
Given multiple observables, $O_i$, that each provide some discrimination on
the two possibilities, either the star is a companion (c) or it is a field
star (ie, not c  or $\neg c$), the standard Bayesian formula in this case is,

\begin{eqnarray} 
P(c\,|\,O_1;O_2;...; O_i) = \frac{\prod\limits_i P(O_i\,|\,c)  P(c)} 
{\prod\limits_i P(O_i\,|\,c) P(c)+\prod\limits_i P(O_i\,|\,\neg c)  P(\neg c)}
\label{eq:Bayes}
\end{eqnarray}

Where the numerator has the product of available probabilities for
each observable having its observed value assuming that the candidate
is a companion, $P(O_i|c)$.  
The numerator also includes a prior
probability term, $P(c)$, in which knowledge of the companion
probability of the ensemble of candidates or additional knowledge of the
companions can be introduced.  
If no prior knowledge is available, then this term can be set
to $\frac{1}{2}$, and at least one will have a rank ordering in the
probabilities of the candidates.  
The denominator has a repeat of the
numerator, plus a similar product of probabilities, except this time
the assumption is that the candidate is not a companion.

For this work, the discriminating observables are radial separations $r$ from parallax, proper motion differences
$\Delta \mu$, 
so the above formula for the posterior probability of being a companion is:

\begin{eqnarray} 
P(c\,|\,r;\Delta\mu) &=& \frac{P(\Delta\mu\,|\,c)P(r\,|\,c)P(c)}
{P(\Delta\mu\,|\,c)P(r\,|\,c)P(c) + (1-P(r\,|\,c)) (1-P(r\,|\,c))(1-P(c)) }
\label{eq:probs}
\end{eqnarray}

With the prior P(c) set to 0.5 this provides probabilities with a starting assumption
that each candidate is just as likely to be a field star as a companion.
One can improve on this by providing the probability of being a
companion based on statistics of the field stars that are nearby in angle,
proper motion, and distance separation $\Delta d$.

\begin{eqnarray}
P(c) = \frac{(1-P(\Delta d\,|\,f))(1-P(\Delta\mu,\theta\,|\,f))P(p)}
{ (1-P(\Delta d\,|\,f))(1-P(\Delta\mu,\theta\,|\,f))P(p)
+ P(\Delta d\,|\,f)P(\Delta\mu,\theta\,|\,f)(1-P(p)) }
\label{eq:prior}
\end{eqnarray}

Here, $P(\Delta d\,|\,f)$ is the probability that one or more field stars
fall with radial distance less than the candidate's distance from the primary.  
Therefore, $(1-P(\Delta d\,|\,f))$ is the probability that no field stars randomly fall 
in this range.
The term $P(\Delta\mu,\theta\,|\,f)$ is the probability that one or more field
stars happens to have proper motion and angular separation more similar 
to the primary than the candidate.

The term $P(p)$ is the probability that the
provisional primary, selected in the manner in which it has, is a primary,
\ie it has at least one wide companion in the radial region that we are
exploring.
It is perhaps somewhat dismaying at first that $P(p)$ is needed to derive
individual probabilities since normally the individual
probabilities would be needed to derive it. 
However, as we show, it is possible to conduct control experiments, 
using the actual catalog data, to constrain $P(p)$.
 
\subsubsection{$P(\Delta\mu\,|\,companion)$ and $P(r\,|
\,companion)$}
\label{P(mu|c)}

To calculate the probability that a binary star would have a given
proper motion difference, one could calculate the intrinsic
probability density function of velocities for random orbits from
Kepler's Laws and take into consideration distances, errors in
distances, and errors in proper motion observations.  The probability
of having some specified observed proper motion difference $\Delta\mu$
assuming it is a companion, is given by the complementary cumulative
distribution function (CCDF) which is the integral of the PDF over
proper motion differences greater than the observed value.
\begin{eqnarray}
P(\Delta{\mu}\, |\, c) &=& \int \!\!\! \int_{\mu>|\Delta{\mu}|} 
PDF(\Delta\vec{\mathbf{\mu}})
dA_{\vec{\mathbf{\mu}}}
\label{eq:p_pm}
\end{eqnarray}

However, it is more straightforward 
to create a simulation of the star catalog (described in
\S~\ref{sec:sim}), add simulated binary orbits, add observational
errors and then form the histogram of the distribution of proper
motions.  Per Eq.~(\ref{eq:p_pm}), the cumulative distribution is
reversed and this provides estimates of the probability of a companion
to have the observed proper motion value.  These probabilities 
are fit sufficiently well by the following form for the
components parallel and perpendicular to the motion of the primary:
\begin{eqnarray}
P(\mu_{\perp}\, |\, c) &=& \exp[{-(\mu_{\perp}/\mu_0)^{\alpha_0}}], \\
P(\Delta \mu_{\parallel}\, |\, c) &=& \exp[{-(\Delta \mu_{\parallel}/\Delta 
\mu_1)^{\alpha_1}}].
\end{eqnarray}
The probability that a companion would have both components greater
than the observed ones is,
\begin{eqnarray}
P(\Delta \mu\, |\, c) &=& 1 - (1-P(\mu_{\perp}\, |\, c))(1-P(\Delta \mu_{\parallel}\, |\, c) 
).
\end{eqnarray}

The parallel and perpendicular coordinates are used here because
our first-order error analysis indicates that most geometric effects
will be parallel to the motion of the primary (see
\S\S\ref{sec:Geometric_ProperMotion_Differences} and
Eq.~(\ref{eq:Delta_mu}) below).

For the probability based on observed 3-d radial separation, we calculate the period {$\cal P$} using $({\cal M}_p+ {\cal M}_c){\cal P}^2 = r^3$ and apply it to the CCDF of the DM91 distribution modified to allow for individual errors in parallax that reflect into errors in $r$ and period of the orbit.  
The CCDF of this log-normal distribution is given by the complementary error function:
\begin{eqnarray}
P(r\,|\,c) = \mathrm{erfc}\left(\frac{\log({\cal P}) - 4.8}{\sqrt{2} \sqrt{2.3^2 + \epsilon_{log\cal P}^2}}\right).
\end{eqnarray}
Of course, for radial separations of several parsecs where the system is no longer bound the DM91 distribution looses meaning, but it continues to be useful in providing a steeply descending function.

The error in $r$ is given by:
\begin{eqnarray}
\epsilon_r^2 =  &[&\left(  (\epsilon_{\pi_p} x_p^2)^2 + (\epsilon_{\pi_c} x_c^2)^2 \right) (x_p - x_c)^2\\
\nonumber &+&  \left(  (\epsilon_{\pi_p} y_p^2)^2 + (\epsilon_{\pi_c} y_c^2)^2 \right) (y_p - y_c)^2\\
\nonumber   &+& \left(  (\epsilon_{\pi_p} z_p^2)^2 + (\epsilon_{\pi_c} z_c^2)^2 \right) (z_p - z_c)^2 ] /r^2
\end{eqnarray}

But, we need the error in the log of the period due to uncertainty in distance,
\begin{eqnarray}
\epsilon_{\cal P} = \frac{3 r^2 }{2 {\cal M P}}\epsilon_r = \frac{3 \cal P }{2r}\epsilon_r \\
\epsilon_{\log{\cal P}} =  \frac{\epsilon_{\cal P}}{{\cal P}\ln(10)} = \frac{3 \epsilon_r}{2\ln(10)r}.
\end{eqnarray}

\subsubsection{$P(\Delta \mu,\theta\,|\,field)$ and $P(\Delta d\,|\,field)$}

We seek the probability that one or more field stars would have a
proper motion as close or closer to the primary as the candidate given
the local density if stars per unit spatial and proper motion area.
The number density is given by:

\begin{eqnarray}
\rho_f &=&  \frac{N_f}{\pi^2(\Delta \mu_{outer}^2 - 
           \Delta \mu_{inner}^2)(\theta_{outer}^2 - \theta_{inner}^2)}.
\end{eqnarray}

The formula for the probability of one or more stars, chosen from a
homogeneous distribution with this density, falling at separation less
than $\theta_{pair}$ and proper motion difference less than $\Delta
\mu$ is:
\begin{eqnarray}
P(\Delta \mu,\theta\, |\, field) 
   &=& 1 - e^{- \pi^2\rho_f  \theta_{pair}^2\Delta \mu^2}
\end{eqnarray}

Even though this formula is formally for a constant density distribution it
works well for a density changing linearly in the coordinates.  
However, near the peak of the density
distribution with proper motion along a line of site, the curvature in the density
profile can cause substantial error in the local density and it is best to
simply avoid this calculation near the peak.  
Since the peak is where the density becomes very high and probabilities are low, this
region needs to be avoided anyway. 
Therefore, we only consider primaries with a field density below a 
given threshold (Table~\ref{tbl:params}).

The probability that one or more field stars would happen to have an
observed distance that is less than the $\Delta d$ observed for a
candidate star is given by similar formulae,
\begin{eqnarray}
\lambda_f &=& \frac{N_f}{2\Delta d_{max} }\\
 P(\Delta d\, |\, field) &=& 1 - e^{-{\lambda_f  \Delta d}}
\end{eqnarray}

\subsection{Geometrically induced proper motion differences.}
\label{sec:Geometric_ProperMotion_Differences}

With increasing stellar separation there is a growth in the proper
motion difference caused by various projection effects, even if
the two stars have the same space motion.  However, in most cases
some compensation can be made for this.  We can first look mathematically at the
gradient of $\vec{\mathbf{\mu}}$ due to spatial separations at
constant space velocity.  
To search for very wide binaries, we start by presuming space velocities 
are essentially the same for the sytem's stars. 
For a binary with a solar mass primary separated by more than 0.01 pc,
the orbital velocities are
$<$ 0.78\ \kms.  Beyond 25 pc, this corresponds to $\la$ 6.6 \masyr\ in
proper motion differences.  
In Galactic coordinates $(\ell,b)$, the vector of
proper motion is composed of the two projections of the 3-d velocity onto
the unit vectors in the longitude and latitude directions, divided by
the distance d to the star.
\begin{eqnarray} 
\hat{\bm \ell} &=& \frac{\hat{\bm z} \times \vec{\bm d}} 
   {\| \hat{\bm z}\times \vec{\bm d} \|} = (-\sin{\ell},\,\cos{\ell},\,0),
   \label{eq:lhat}\\
\hat{\bm b} &=& \frac{ \vec{\bm d} \times \hat{\bm \ell}}
   {\| \vec{\bm d} \times \hat{\bm \ell}\|} =
   (-\sin{\gb}\cos{\ell},\,-\sin{b}\sin{\ell},\,\cos{\gb}), 
   \label{eq:bhat}\\
\muvec &=& (\mu_{\ell},\mu_{b}) = 
     (\frac{\vec{\bm{v}} \cdot \hat{\bm \ell}} {d},
   \, \frac{\vec{\bm v}  \cdot \hat{\bm \gb   }} {d}),
   \label{eq:muvec}\\
\mu_\ell &=& \,\, \left(
   -v_x \sin{\ell} + v_y\cos{\ell}\right)/ d
   \label{eq:mu_l}\\
\mu_b &=& \left( \left(
   -v_x \cos{\ell} - v_y\sin{\ell}\right)\sin{\gb} + v_z \cos{\gb} \right) 
   / d
   \label{eq:mu_b}\\
v_r   &=&  \,\, \left(
   +v_x \cos{\ell} + v_y\sin{\ell}\right)\cos{\gb} + v_z \sin{\gb} \,\, ,
   \label{eq:vr}
\end{eqnarray}
where $\hat{\bm z}$ is the unit vector to the North Galactic Pole.  
If distances are in pc and proper motions are in \asyr, then the radial 
velocities are in AU yr$^{-1}$ (1 AU yr$^{-1} \simeq 4.74\ \kms$ ).
The velocity $\vec{\bm v}$ is relative to the sun so it is formally 
$\vec{\bm{v}}_* - \vec{\bm v}_{\Sun}$, where $\vec{\bm v}_{\Sun}$ is the
velocity of the sun in the Local Standard of Rest (LSR).

The derivatives of the direction vectors:
\begin{eqnarray}
\frac{\partial \hat{\bm \ell}  } {\partial{\ell}} &=& (-\cos{l},
\,-\sin{l},0),
\qquad{\frac{\partial{\hat{\bm \ell}} } {\partial{\gb}}} = (0,0,0),
\label{eq:dlhat}\\
\frac{\partial \hat{\bm \gb}} {\partial{l}} &=& -\hat{\ell} \sin{\gb} ,
\qquad{\frac{\partial{ \hat{b}}} {\partial \gb}} = -\hat{\bm d},
\label{eq:dbhat}
\end{eqnarray}
 can be used to derive the gradients in spherical coordinates to get a first
order assessment of the geometric induced variations on the difference
in proper motion between two stars moving at the same  space
velocity:

\begin{eqnarray}
\Delta \vec{\bm{\mu}} &=& 
- \vec{\bm{\mu}} \frac{\Delta d} {d} 
+ \bigg(\mu_\gb  \sin{\gb} - \frac{v_r} {d} \cos{\gb},\, 
- \mu_{\ell} \sin{b} \bigg) \Delta \gl 
+ \bigg(0,-\frac{v_r} {d}\bigg) \Delta {\it b} \, \, ,
   \label{eq:Delta_mu}
\end{eqnarray} 
where $\Delta \ell$ and $\Delta b$ are the difference in longitude and
latitude, in radians.
Similarly, we can look at the first derivatives of the radial velocity
which, if known for both components, can be taken into account when
assessing whether the pair is really co-moving:
\begin{eqnarray}
\Delta v_r &=&  
   d \left( \mu_{\ell} \cos{b} \Delta \ell + \mu_{b} \Delta b \right)
   \label{eq:Delta_V_R_apx}
\end{eqnarray}
As an example, let us examine a case where $b=45\ad$, $\mu_\ell=\mu_b=0.2$
 \asyr, the radial velocity is -20 \kms, the primary's
  distance is 20 pc, the companion is 1 pc farther away than the primary, 
  and they are separated by 1 degree per coordinate ($\Delta \ell =
  \Delta b = 0.0175$, separation $\apx\onehalf$ pc ). We then have:
\begin{eqnarray}
\Delta \mu_\ell &=& 
   -0.2 \cdot 1/20 + 
    \frac{\sqrt{2}}{2} \left( 0.2 + 20/(4.74\cdot 20) \right) \cdot 0.0175
    = -0.02\, \asyr = -20\, \masyr\nonumber\\
\Delta \mu_b &=&
   -0.2 \cdot 1/20 + \left( -\frac{\sqrt{2}}{2} \cdot 0.2 + 
                            20/(4.74\cdot 20) \right) \cdot 0.0175
    = -9\, \masyr \nonumber\\
\Delta v_r &=& 4.74 \cdot 20 \cdot 0.2 
\left( \frac{\sqrt{2}}{2} + 1 \right) 
    \cdot 0.0175 = 0.6 \, \kms \,\, .\nonumber
\end{eqnarray}
Note that the changes in proper motions can be quite significant
with respect to the measurement errors (typically 1 -- 3 \masyr\ and 1 -- 3
\kms).   
This implies that very wide binaries of several degrees in separation are,
 in general, 
not found by commonality in  proper motions,  unless proper geometric corrections have been applied.  

Eq.~\ref{eq:Delta_mu} implies that when relative distances are unknown, one 
can still make useful corrections using the proper motions and radial 
velocities.  And, when radial velocities are also unknown one can still make 
useful corrections, by assuming $v_r=0$, that become highly accurate for 
$\Delta \mu_\ell$ near the poles.

Also, one can mitigate somewhat against large uncertainty in the
$\Delta d$ term in Eq.~(\ref{eq:Delta_mu}), by taking note that
this term is parallel to a system's overall proper motion.  
Therefore, one should   treat the parallel and
perpendicular components of the proper motion separately to take
advantage of the 
perpendicular component which needs no $\Delta d$ correction term and hence is less noisy.

The first order correction, Eqs.~(\ref{eq:Delta_mu}), is typically good
to $\lesssim 10$\% for angles $<$ 5\degree\ and distance differences of $<30\%$, 
except near $b = \arctan (v_r/d\mu_b)$ where the $\Delta \mu_l$ term goes to zero for changes in angle.  
To explore out to separations of $\sim$10\degree\  
and reach several pc physical separations,
a full (nonlinear) correction for space velocities is made.   
We calculate the 3-d space velocity of each primary
from its proper motion and radial velocity (Eqs.~\ref{eq:v_x} --
\ref{eq:v_z}), if we have it,
and, using Eqs.~(\ref{eq:mu_l}) -- (\ref{eq:mu_b}), calculate the 
proper motion that this space motion implies, {\it if it were in the
direction of each companion candidate}. 
If there is no radial velocity for the primary, then the system is 
assumed to be at rest in the LSR. 
In those cases, the projection of the 
the Solar Motion in the radial direction is subtracted,
using (10.0, 5.25, 7.17) \kms\ for the Solar Motion.
As described in \S~\ref{sec:Radial_Velocities}, for $d < 100$ pc, there are
radial velocity measures for most of the primaries.

The conversion to space velocities  is the inverse of
Eqs.~(\ref{eq:mu_l}) -- (\ref{eq:vr}) and has solution:
\begin{eqnarray} 
v_x &=& v_r \cos{b} \cos{\ell}  - d \mu_\ell \sin{\ell} - d \mu_b \sin{b} 
\cos{\ell} 
\label{eq:v_x}\\
v_y &=& v_r \cos{b} \sin{\ell}  + d \mu_\ell \cos{\ell} - d \mu_b \sin{b} 
\sin{\ell} 
\label{eq:v_y}\\
v_z &=& v_r \sin{b}  +  d \mu_b \cos{b}
\label{eq:v_z}
\end{eqnarray}
  
To avoid excessive error arising from distance uncertainties at distances beyond
25 pc, we assume the companion is at the distance of the primary.

\subsection{Simulation}\label{sec:sim}

A program for creating simulations of the Solar Neighborhood
distribution of stars with binary systems was written to test
the methodology, search for optimal parameters, and to understand the origin 
and behavior of false positives.  
The simulations cover a large range of parameters to understand the 
reliability and broadness of applicability of the methodology.  
The simulations also are used as a quick way to derive the shape of the cumulative distribution
functions for the observables in an ensemble of random binary systems. 

In the simulations, stars are set down with a distribution of positions
and velocities that statistically imitate the
HIP2 after observational errors are included.
Observational errors representative of  moderately faint stars in TY2 are used:
 $\epsilon_\mu =$ 1.5 \masyr, $\epsilon_{\ell} = \epsilon_{\it b} =$ 1 \mas. 
For parallax errors of the simulated stars, we use
$\epsilon_\pi =$ 1 \mas\ to represent the HIP2 data.
Masses are distributed with the power law $N \sim M^{-2.3}$ from 0.8 to 15 $\Msun$ and luminosities 
follow a mass-luminosity relation, $L \sim M^{-3.5}$.
An arbitrary fraction of the stars are given companions.
The companions do not themselves
host companions, \eg no hierarchical systems are created.  
The orbital elements: eccentricity, inclination, longitude of the ascending node,
longitude of the periapsis, and epoch are
random uniform distributions.  
The distribution of periods is chosen to be log-normal without a long period cutoff.    

\begin{figure}
\epsscale{1.0}
\plottwo{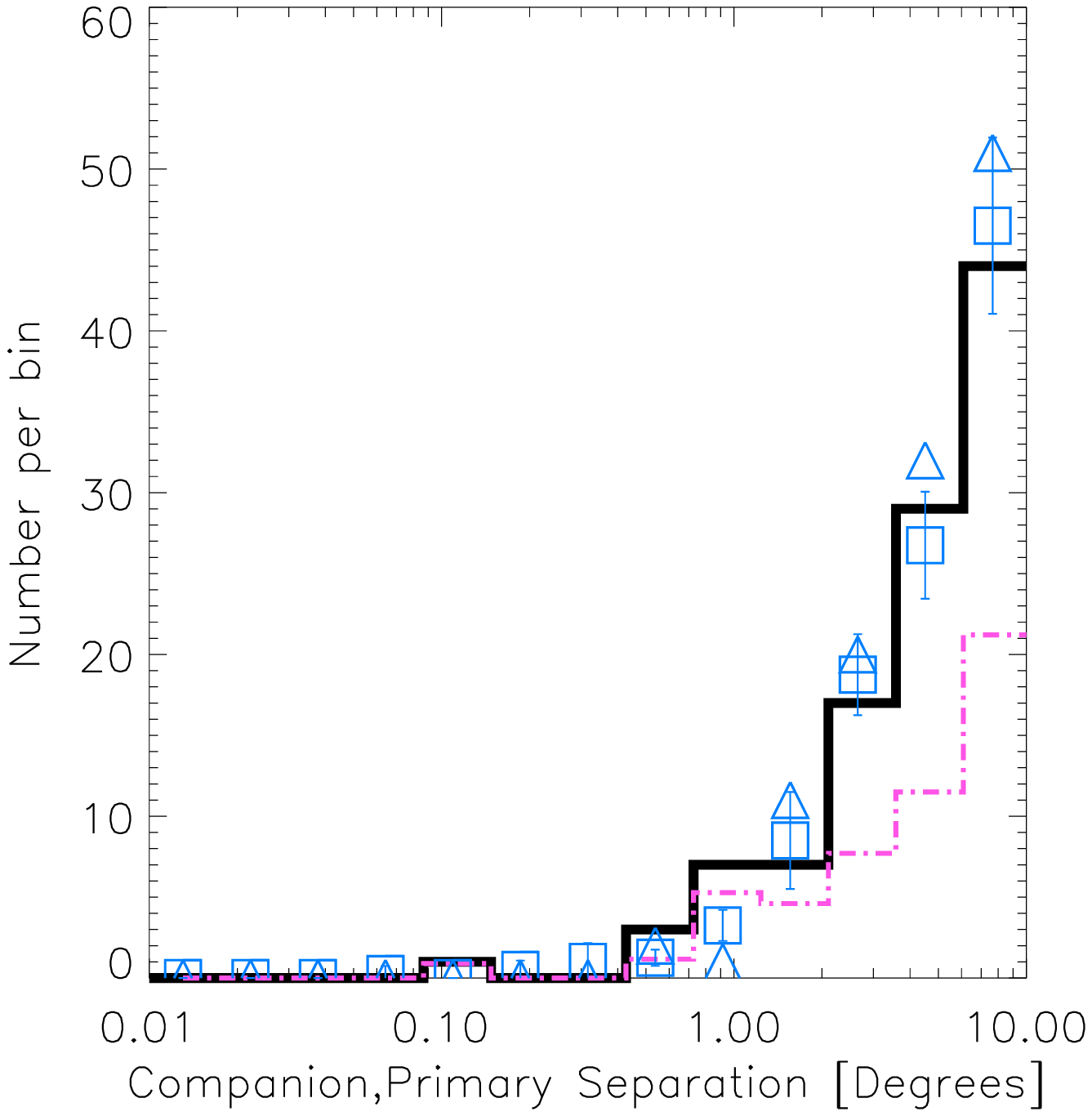}{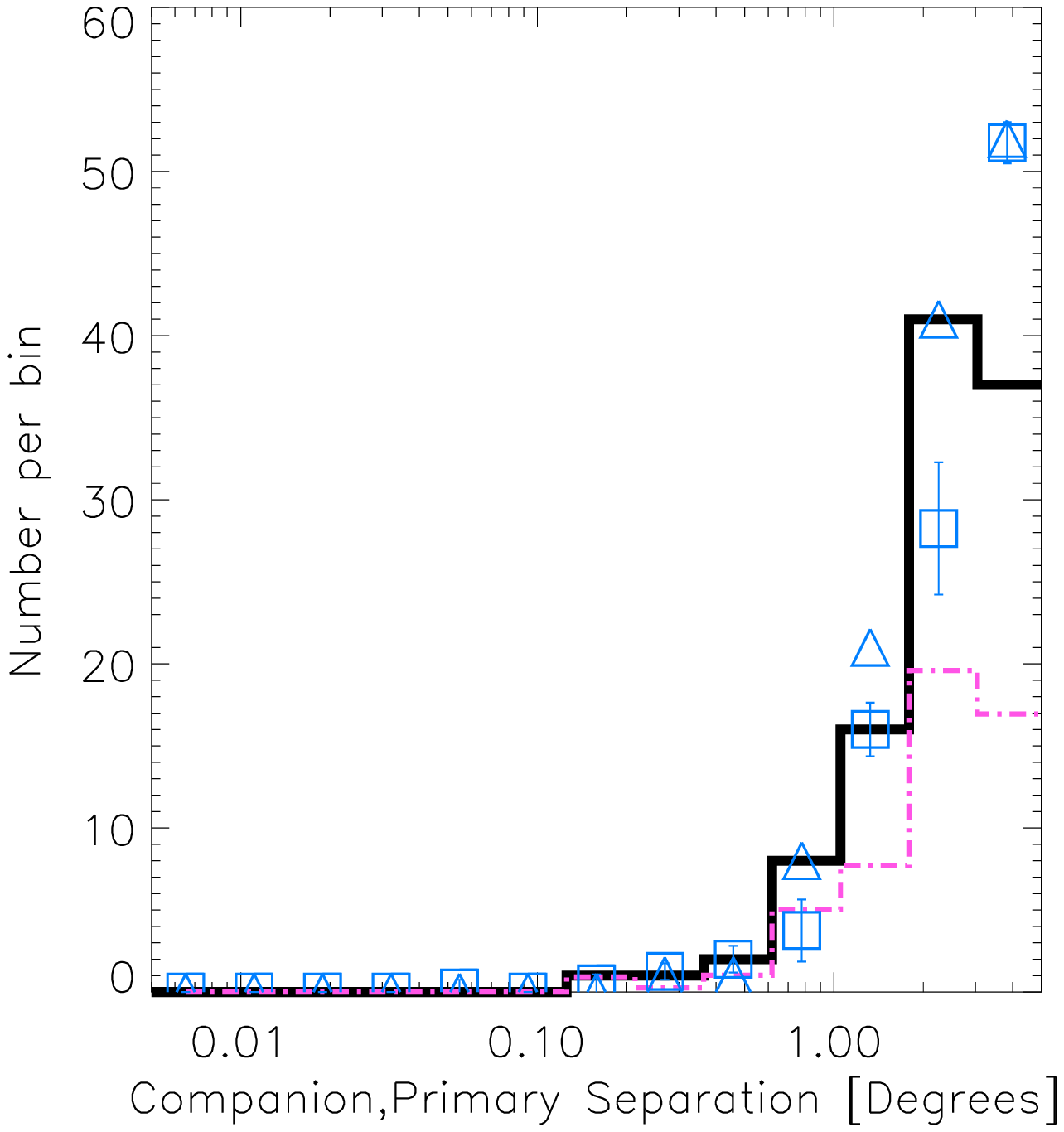}
\caption{
 {\it Simulation: Number vs. Angular Separation with No Companions}
  - The Local Neighborhood is simulated but with no binary systems included.   
The diagrams show numbers of companions per logarithmic separation bin with  
probability $>10$\%.  Since there are no companions here, 
these are all false positives.
The left side shows results for primaries between 25 -- 50 pc, 
and the right side is for 50 -- 100 pc. 
The solid thick line gives the numbers found in an analysis of the 
simulation similar to one we do for the Hipparcos Catalogue.
The control experiment in which {\it b} = -{\it b} for the primaries is shown as 
triangles.
The average over 4 ```rethrow" control  experiments is shown as blue
squares and the variance is shown in error bars.  
The purple dash-dotted line is the sum of the probabilities of 
companions within each separation bin. 
\label{fig:sim_noprimes}}
\end{figure}

We first present the relatively small false positive rate for a simulation in 
which no binary stars were generated.  
In Fig.~\ref{fig:sim_noprimes} the numbers of companions with probabilities $>0.1$ vs.
angular separation are shown  (black thick solid line), 
where the associated primaries are in the 25 -- 50 pc (left) and
50 -- 100 pc (right) distance ranges .  
The ``false positive" rate is  kept  low  by using a low value for the prior
$P(p)$, namely 0.20 for 25 -- 50 pc and 0.07 for 50 -- 100 pc. 
These values for $P(p)$ are chosen because they work reasonably well for all that follows.  

Fig.~\ref{fig:sim_Deg} shows results of an analysis of a 
simulation in which the semi-major axes are set by the DM91 distribution of periods 
but using only periods longer than the peak of the distribution at 173 yr.
About 21\% of the stars are primaries and 28\% are companions.
These rates are much higher than the observed one, considering that the mass
function only goes down to 0.8 $\Msun$,
but it provides many companions and much confusion to better test the procedure.
The primaries have 1, 2, 3, or 4 companions with frequency of roughly 67.8, 
27.4, 4.5, and 0.2\%, respectively.  
The green dotted line shows the number of companions created per separation bin.
The solid thick line shows the number of companions found with probabilities
$>10$\%.  
The green thin solid line gives the number of correct primary-companion 
associations found.
The lower green dashed line gives the number of companions ascribed to be
primary-companion pairs, but neither was
an input primary, \ie it gives the number of secondary-tertiaries pairs, etc. found. 
The procedure recovers better than 90\% of the multiple systems for 
separations out to 2\degree\ for
 25 -- 50 pc and 1\degree\ for 50 -- 100 pc or about a parsec.

In Fig.~\ref{fig:sim_Deg_100} , the period distribution for the simulation is multiplied
by 100 to shift the distribution to larger semi-major axes.  Now, one can
see that there is still reasonable recovery ($>$20\%) of companions even at
separations of 10\degree\ for 25 -- 50 pc region and 5\degree\ for 50 -- 100 pc.  For 0 -- 25 pc,
we get good fractional recovery until 20\degree. 

The resulting distributions of observables for this simulation are used 
to set the coefficients presented in Table~\ref{tbl:params}.  
These coefficients do not depend on the fraction of stars that hold binaries, 
nor do they depend strongly on the shape of the separation distribution.  
They do depend fairly strongly on the assumed observational errors which essentially 
determine the widths of the cumulative distributions.  
However, as the width of the distributions change, the value of $P(p)$ needs to adjust 
in a direction to keep the sum of the probabilities constant near the values
indicated by the control experiments (next section).  
The net result is that the assigned probabilities 
change rather slowly with the observational errors assumed. 

\subsection{Control Experiments}

We implement two kinds of control experiments that use the observed data 
directly
rather than the simulation to assess the rate of false positives and
thereby determine the prior probability $P(p)$.  Using the real
data for control tests provides greater confidence that minor
differences between the simulation and reality do not cause
inconsistencies.  In the first method, the negative of the
Galactic latitude is used for each star while it is considered as a primary.
This moves the primary away from its companions and places it in a
region with similar stellar density and velocity field.  Any
stars assigned high probabilities are statistical coincidences or false 
positives.
Although there is asymetry between the two hemispheres in the observed 
proper motion distribution, much of this goes away after transforming to 
the LSR and the remaining should have a small affect on the numbers of 
false positives.

In the second method, all candidates are removed from each primary,
and then each field star is randomly ``rethrown" with new uniform
random 2-d positions within the separation angle limit, $\theta_{outer}$,
and random values of $\Delta \mu$ within a circle in coordinates
($\Delta \mu_{\shortparallel}, \mu_{\perp}$) whose size is set to maintain
the density in phase-space.  The field star distances and brightnesses
are maintained.
When the control experiments are run on a simulation with zero
binaries, the two control methods and the direct analysis of the simulation 
returns approximately the same number of false positives, 
within each probability bin, as they should.
The first method of control experiment, with the reverse sign of $\gb$ 
value of each primary, is shown as blue triangles
in Fig.~\ref{fig:sim_noprimes},  where no binary stars are generated in the simulation.
The second method, with field star rethrown, is shown as blue squares with 
error  bars at the average and rms deviations of 4 realizations.
For the other figures showing number versus separation,  both types of 
control experiments are averaged together  and shown as blue squares.

\section{Application to the Hipparcos Catalogue}

As a first application of our Bayesian probability estimator,
Eqs.~(\ref{eq:probs}) and (\ref{eq:prior}), we examine stars in the
HIP2 brighter than 10th mag in V-band for possible
companion stars brighter than V=12.  
Provisional primaries are separated into three distance
intervals, by parallax distance. There are 1,041 potential primaries
(V$<$10) within 25 pc, 4,152 in the 25 -- 50 pc shell,
and 14,064 in the 50 -- 100 pc shell.
A 15\degree\ radius around the Hyades is cut out when working with the 25 -- 50 pc 
primaries.
A 200\am radius around the Pleiades Cluster and also
around the Coma Star Cluster are cut out for 50 -- 100 pc primaries.
Stars with $\pi<5$ mas are presumed to be too far away and are also
removed.  For companions, about 25,000 stars are brighter than V=12
and are within 110 pc. 

Candidate companions and field stars are selected if their separation
from the primary is 
36\as $< \theta_{lim}<20\degree$ for $d_p < 25$ pc,
36\as $< \theta_{lim}<10\degree$ for $ 25 < d_p < 50$ pc, and
18\as $< \theta_{lim}<5\degree$ for $50< d_p < 100$ pc,
(Table~\ref{tbl:params}). 
Most HIP binaries closer than 36\as\ and within 50 pc would already be known
and their proper motion differences may be substantially
affected by orbital motion, while our methodology is optimized for
the case of low orbital speeds.  
Companions are not constrained to come
from the same distance interval as the primary star.  
Table~\ref{tbl:attrition} presents how many potential primaries there were 
in each distance range and the numbers remaining after dropping ones with no 
stars nearby, then no candidates, then too high of a field density, and finally 
presents the number of primaries and candidates with probabilities over 0.1.  
The total probabilities given for all candidates in the entire separation range 
and for just those with separations $< 1$ pc.

Table~\ref{tbl:params} provides the final set of parameters that are
used in the analysis of each distance intervals and in creating the
tables and figures in this section.  In addition to the coefficients
for the cumulative distributions in each of the observables and the
cutoffs in angle and proper motion for candidates and for field star
counts, the table includes the maximum number of field stars accepted.
Most field stars along a given direction are concentrated in a small
range of proper-motion: consistent with expectation from the Galactic
rotation and the projection of the Solar Motion along the
line-of-sight.  If the phase-space of the star is well centered in
this ``cloud", usually, the probabilities for any companions would
naturally be low and the rate of false positives unacceptably high.  
To avoid this a star is dropped if the field
phase-space density is in the 90-percentile in the density
distribution.  However, in the 0 -- 25 pc region, the number of
candidates is always small, so, it is unnecessary to include this
criterion there. 

Histograms of the separation distributions for the HIP catalog
are shown in Fig.~\ref{fig:hip_Deg} (in degrees) and
Fig.~\ref{fig:hip_pc} (in parsecs).
In each diagram, the black thick line is the
histogram of companions with $P>0.1$, the blue
square symbols are the averages over 5 control experiments, and the red 
line shows the numbers found minus the numbers in the control experiments,
providing an estimate of the range of real physical companions (red error bars).  
This choice for $P(p)$ is simply the value that brings the sum of the 
probabilities in each bin into agreement with 
the number of companions found minus the number in the control experiments.
The sum of the probabilities of companions within each separation bin  
(purple dash-dotted line) has been adjusted by varying $P(p)$,
settling at a value of 0.20 for 25 -- 50 pc and 0.07 for 50 -- 100 pc.  

The degree of agreement is startling to the authors.  
Since the actual distribution at these separations is very different 
from the DM91 distribution used in $P(r|c)$,
one might worry that it would not work at all.  
However, all that is required
for this probability is a function that falls off rapidly enough to
sufficiently suppress the false positives, and the DM91 law happens to work.

For separations up to $\approx 1$ pc the number of false positives
found in the control sample is quite small implying that the
companions found in the real sample are reliable.  
The breakdown by probability interval at various separations is presented in
Fig.~\ref{fig:pseps}; the different colored regions show the
distribution contoured at 0.1, 0.25, 0.75, and 0.95 probability levels.

\subsection{Radial Velocities}
\label{sec:Radial_Velocities}
Radial velocities differences can be used, where available, as a check on the 
reasonableness of the probability assignments; therefore we searched the 
literature for radial velocity measures of HIP stars.
Because some radial velocity (RV) catalogs in CDS\footnote{
  CDS: Centre de Donn{\' e}es astronomiques de Strasbourg;
  http://cds.u-strasbg.fr/}       
are not incorporated when ``querying by identifier'' or ``querying by
coordinate'' from SIMBAD, we decided to use SIMBAD plus extract RV catalogs 
from the CDS archives. 
A given HIP star may be found in several catalogs and 
therefore have several values.  
The data is taken from the six catalogs enumerated below.  
The order is in increasing reliability and thus 
increasing priority, \eg if a later catalog provides a velocity, we use 
that one.  
These numbers of obtained radial velocities with distance are plotted in 
Fig.~\ref{fig:vr} and tabulated in 
Table~\ref{tbl:25pc}, \ref{tbl:50pc}, \ref{tbl:100pc}.

\begin{itemize}
\item The SIMBAD data base. While we did extract the radial velocities
  in batch mode, the errors could not be obtained in that way. So 
  all RV errors for these stars are set to 10 \kms. We find 36,884 HIP stars
  with RV data in SIMBAD.
\item {\it The General Catalogue of Mean Radial Velocities} [GCRV;
  \citet{GCRV_2000}; VizieR cat. III/213]; following the description
  in VizieR catalog III/21 \citep{GCRV_1953}, the following
  values for the RV errors based on the quality factors are assigned: quality=A
  $\rightarrow \epsilon_{RV}$=0.5 \kms,
  q=B $\rightarrow \epsilon_{RV}$=1.2 \kms, q=C $\rightarrow
  \epsilon_{RV}$=2.5 \kms, q=D $\rightarrow \epsilon_{RV}$=5.0 \kms.  Stars 
  with quality E ($\epsilon_{RV} \ge 20$ \kms) are excluded.
  These errors correspond more or less to the midpoints of
  the ranges specified by \citet{GCRV_1953}. We find 21,120 HIP stars
  in the GCRV.
\item {\it The Bibliographic Catalogue Of Radial Velocities} [BCRV;
\citet{BCRV_2000}; VizieR cat. III/249], which is up to date till
2006. This catalog required significant attention as it does not
list errors on the RVs, while also many stellar names do not conform
to the current SIMBAD convention. This may not be too surprising
since \citet{BCRV_2000} compiled RV data from almost 1300 different 
publications. However, the vast majority of stars
are found in just 33 different publications. We read
those 33 publications and estimated an RV error for each of them. Stars 
from other publications are, somewhat
arbitrarily assigned RV errors of 10 \kms.  1,178 stars are found in
more than one publication, and their weighted average values
and errors are used in our database. In total, we find 14,279
HIP stars in the BCRV that are {\it not} in the GSCN catalog, described below.
   \item {\it The Catalogue of Radial Velocities of Galactic Stars with
     Astrometric Data, the Second Version} [CRVAD; \citet{CRVAD_2007}; 
     VizieR cat.
     III/254]. The same error assignment is used here as for the GCRV
     above, and stars with RV quality=E are not used.  We
     find 41,740 HIP stars in the CRVAD.
   \item {\it The Geneva-Copenhagen Survey of the Solar Neighborhood}
     [GCSN; \citet{GCSN}; VizieR cat. V/117\footnote{The original
       catalog V/117 is hard to find on VizieR because it is claimed
       to be obsoleted by V/130. However, V/130 contains significantly
       less information, \ie neither mass estimates nor the raw RV
       information. This catalog can be accessed by going directly to
       the source:
       \url{http://vizier.u-strasbg.fr/viz-bin/VizieR?-source=V/117}.}].
     The ``median errors'' as listed in the GCSN are used. We find
     11,900 HIP stars in the GCSN that were not taken from the GCRV.
   \item {\it The Radial Velocity Experiment: Second Data Release} [RAVE;
     \citet{RAVE_2nd_2008}; VizieR cat. III/257]. We use the errors as
     presented in the RAVE catalog. If more than one entry is 
     present per HIP star, the weighted average for both
     the value and the error are used. We find 393 HIP stars in the RAVE data
     set.
\end{itemize}
Altogether, we have 43,047 radial velocities out of 113,942
HIP stars with positive parallaxes: the average completeness is 37.8\%.
The additional catalogs added 6,161 RVs to that available in SIMBAD alone.  
The completeness fraction is a strong function of distance 
(Fig.~\ref{fig:RV_completeness}) and apparent magnitude.  
About 48\% of
stars at a distance of 100 pc have a measured radial velocity, but
within 100 pc, RV data is available for almost two-thirds of stars.
Of the systems where both
the primary and candidate have a measured RV, \apx87\% systems have
errors $<$10 \kms, with an average of about 1.5 \kms.

 For those stars with RV data in multiple catalogs, we compared the
  various values to assess their external accurcy.
  If a star is in fact a
  binary, then the cataloged values might have been taken at different
  orbital phases, which would result in a scatter that is larger
  than expected on the basis of the reported internal
  errors. We kept track of 
  the range of the reported RVs for a given star, and the errors.
  If this range exceeds 1.6 times the error {\it and} the
  velocity range exceeds 9 \kms, then the star is deemed to have a
  discrepant RV, which may be the result of (unsuspected) binarity.
  About 13\% of HIP stars are classified as some sort of
  binary, while the ones with discrepant radial velocities are classified as
  binaries about four times more often (\apx55\%). For
  our subsample of candidate very wide binaries with discrepant
  RVs, about 75\% are known or suspected binaries. The whole
  sample of very wide binary candidates has a rate of known/suspected
  binaries that is 3x larger than for the whole HIP. 
  It is important
  to note that these known/suspected binaries most often refer to
  companions closer to our candidates, {\it not} the
  candidates we report in this paper. 
We interpret this as evidence that very
wide binaries are often found in hierarchical systems, as suggested by a
  number of authors \citep{MZH_2008, Caballero_2009, Caballero_2010,
    Kouw_WideBinForm_2010}.

\subsection{Stellar Masses}
\label{sec:Stellar_Masses}

It's useful to obtain mass estimates for the stars in these systems to learn 
how their separations compare with their nominal tidal radii. 
Absolute magnitudes ($M_V$) and
stellar masses  (${\cal M}$) are assigned based solely on stellar color (B-V),
 via the mass-luminosity-color relation for the main-sequence (MS).  
We note that the so-determined $M_V(B-V)$ relation forms a lower envelope to 
the HIP color-magnitude diagram for main-sequence stars: our relation is
  close to, but not identical to the zero-age main sequence (ZAMS)
  $M_V(B-V)$ relation.
The starting point is Tables~3.13 and
3.10 in \cite{BM_1998}, which list stellar colors, masses and absolute
magnitudes as a function of spectral type. Next,
the colors are updated in the following way. 
1) The BVRI colors for O5
-- M5 stars are taken from \citet{Cox_Allen}, where this
source is also used for the values of effective temperature ($T_{eff}$). 
2) BVRI colors are preferentially taken from \citet{Bessell_1990} for types 
M0 -- M6. 
3) The previous references are superseded by VRIJHK data from 
\citet{Bessell_1991} for
M6 -- M7.5 dwarfs.
4) Average B-V colors are computed for
late-type M dwarfs by averaging the colors and $M_V$ of several such dwarfs 
extracted from the NSTARS
database\footnote{\label{foot':NSTARS} We estimate B-V=1.91 \pmt\ 0.064, 1.99 
\pmt\ 0.009, 2.05 \pmt\ 0.078, 2.16 \pmt\ 0.8 and 2.10 for
types M5.5, M6.0, M6.5, M7.0 and M8.0, respectively.}.  
5) We examine the B-V data for early M dwarfs from
\citet{Koen_2010}, and , to be consistent with the upper MS, we determine a
lower-envelope to their HR diagram\footnote{We use: B-V=1.47, 1.51, 1.577,
  1.677 for types M1, M2, M3 and M4, respectively.}.

For all cases, we use the dependence of a given color on $T_{eff}$ to
interpolate over missing values\footnote{Not all stars in HIP have
  B-V colors.  Furthermore, because the B-V colors in HIP are derived
  in a non-homogeneous manner (partly derived from ground-based
  observations and partly from the TY1 photometry), we use our own
  estimation for the B-V values on the Johnson system based on TY2
  colors: (B-V)$_J$ = 0.85\,(B-V)$_{T2}$. This transformation is
  accurate to \pmt\ 0.071 mag, which is 2.2 times larger than the
  errors on B-V as listed in HIP. Note that we use the TY2 colors,
  which differ substantially (at fainter magnitudes) from the TY1
  colors listed in HIP.}.
Then, for each star, their B-V values are used to compute their
  main-sequence masses. However, still not all stars have B-V values,
  and we estimate their masses from the weighted average of a number
  of other color-mass relations: ${\cal M}_{V-I}$, ${\cal M}_{V-J}$,
  ${\cal M}_{V-H}$, ${\cal M}_{V-K}$, ${\cal M}_{J-H}$, ${\cal
    M}_{J-K}$, ${\cal M}_{H-K}$ and 
    ${\cal M}_{H-K}$\footnote{The J, H \& K magnitudes (and errors) are 
    extracted from the 2MASS
  database.}.

Lastly, a correction is made for the effects of stellar
  evolution as stars evolve off the zero-age main sequence.
  The GCSN catalog
  \citep{GCSN} provides stellar masses corrected for stellar
  evolutionary effects, absolute magnitudes and $(B-V)$ colors for
  14,955 HIP stars.  We compute a two-dimensional look-up table (map),
  with the average mass as a function of (B-V) and $M_V$.  
  This contour map, Fig.~\ref{fig:Stellar_Masses}, shows for any star that
  falls in a defined part of the map, the corresponding mass value used. 
  For stars in undefined parts of the map, 
  the color-mass relation for the MS is used.
  The masses are listed in column \#14 of tables
  \ref{tbl:25pc} -- \ref{tbl:100pc}.  
  Note that these masses
  are insufficient to decide whether or not a potential pair is
  bound because, quite often, the system contains stars that are not a
  part of HIP.

\subsection{Tabulated Results}
\label{sec:Results}

The focus of this research has been on very wide companions in 
the 25 -- 100 pc interval; however, we 
have applied a similar methodology to the 0 -- 25 pc interval even though 
it has not been optimized for this regime.  
Fortunately, a number of high probability companions are discovered in 
this region out to 20\degree\  in separation.
The physical systems and their probabilities are presented in tables
\ref{tbl:25pc}, \ref{tbl:50pc} and \ref{tbl:100pc} for the distance
ranges 0 -- 25 pc, 25 -- 50 pc and 50 -- 100 pc, respectively.  
The names (columns \#1 \& \#2), positions (cols. \#3 \& \#4) and the
visual magnitude (col. \#5) are taken from HIP. The spectral type
(\#6) is taken from SIMBAD\footnote{\label{foot:SIMBAD}
 http://simbad.u-strasbg.fr/simbad/}. 
We also list in this column one of ten possible types
on the basis of the 67 ``Other object type'' identifications in SIMBAD
(such as, ``*in**,'' ``EB,'' ``YSO,'' mean: ``star in double star,'' 
``Eclipsing binary'' and ``young stellar object,'' respectively). 
The codes are: UkN (not known), SpB (spectroscopic binary), 
EcB (eclipsing binary), BiN (other binary), RoT (star with high rotation velocity), 
VaR (variable star), YnG (young/pre-MS star), ClN (star in cluster of nebula), 
F/E (flare or eruptive star), OtH (generic star or other object). 

The proper motions (\#7 \& \#8) are from TY2 when available, 
otherwise they are from HIP2.  
Columns \#9 and \#10 list the proper motions differences {\it corrected} for
geometric effects as described in
\S\ref{sec:Geometric_ProperMotion_Differences} above. 
In detail, our procedure is as follows. First, for each
potential companion star, its space velocities $(v_x,v_y,v_z)$ is computed
according to Eqs.~(\ref{eq:v_x} -- \ref{eq:v_z}) using their angular 
coordinates, distances, and proper motions. However, beyond 25 pc, 
the distance of the primaries is used because $P(\Delta \mu\,|\,c)$
is meant to give the probability of having $\Delta \mu$ assuming 
that it is a companion, and should not be reduced by a large distance error.
Beyond \apx25 pc, the typical parallax errors,
imply distance errors that significantly
exceed the orbit size and would artificially inflate the
inferred proper motion corrections.
We use the RV of the primary because it is usually the better studied and, 
therefore, is more likely to provide the barycentric velocity of the system.
   
In the next step, these space velocities are transformed into
the expected proper motions if it were at the position of the primary.  
The corrected proper motion
differences listed in columns \#9 and \#10  are the difference between 
the proper motion components of the primary and the candidate if it were seen
with the same projections onto the sky as the primary.
In the same columns,  the errors on
these $\Delta \mu^{cor}$ values are listed, which are derived via propagating
the errors on the observables. The procedure inflates the observed
proper motion errors, typically by factors of 2 to 4. This is to
be expected because the corrected proper motion difference
contains 3 or 4 terms (\cf\ Eq.~[\ref{eq:Delta_mu}]) 
that are RSS-ed together to yield the errors. 
    
The distance (\#11) is the inverse of $\pi$ obtained from HIP2. 
Column \#12 contains the radial velocities compiled in 
\S \ref{sec:Radial_Velocities}.
For Column \#13, the proper motion and radial velocity of the companion are
transformed to LSR space velocities at its position and then the radial 
velocity is calculated from that space velocity translated to the position 
of the primary. 
Beyond 25 pc, we again use the primary's distance in the first step 
instead of the companion's to minimize correction errors due to 
distance uncertainties. 
The errors on the corrected RVs are nearly the same as the errors on the 
measurements themselves because the corrections are typically small.
The RV differences are plotted in Fig.~\ref{fig:vr}.
The separation (\#15) follows from the positions of the
primaries and companions and the distance of the primary. 

Column \#16 gives the probability of the candidate being a true companion 
of the primary according to Eq.~(\ref{eq:probs}).
  
The last column (\#17) provides Bayer-Flamsteed (BF) designations and
common names for the stars.  These are extracted from the
``HD-DM-GC-HR-HIP-Bayer-Flamsteed Cross Index'' \citep{Kostjuk_2004},
with the following modifications: 1) a common name is not listed if
the same name is used for another HD star, unless the system
is a close pair, 2) if more
than one common name was specified, the shortest one is used.
Note that in some cases,
\citet{Kostjuk_2004} uses as the BF designation {\it both} the 
numeric and the Greek designation.
If the star has another HIP star as a companion (as listed in HIP), we add the
known HIP numbers after the ``kn:'' designation. 
  
\subsection{Some Notable Companion Pairs}\label{sec:examples}

We find these unnoticed naked-eye companions ($<$6\uth mag): 
Capella \& 50 Per, $\delta$ Vel \& HIP~43797, 
Alioth ($\epsilon$~UMa),Megrez ($\delta$~UMa) \& Alcor, 
$\gamma$ \& $\tau$~Cen, $\phi$~Eri \& $\eta$~Hor, 62 \& 63~Cnc, 
$\gamma$ \& $\tau$~Per, $\zeta$  \& $\delta$~Hya, 
$\beta^{01}$, $\beta^{02}$ \& $\beta^{03}$~Tuc, N~Vel \& HIP~47479, 
HIP~98174 \& HIP~97646, 44 \& 58 Oph, s Eri \& HIP~14913,
and $\pi$ \& $\rho$ Cep.  
High probabality fainter companions ($>$ 6\uth mag) of stars $V<4$ 
are found for:
Fomalhaut ($\alpha$~PsA), $\gamma$~UMa, $\alpha$~Lib, Alvahet ($\iota$~Cephi), $\delta$~Ara, Chow ($\beta$~Ser), $\iota$~Peg, $\beta$ Pic, $\kappa$~Phe.  and
$\gamma$~Tuc

\subsubsection{The Capella System}
We identify the T~Tauri star 50~Per (HIP~19335) as
a P = 0.2 candidate companion of Capella. 
While these stars have almost
equal radial velocities and corrected proper motion differences,
they are separated by almost 15\ad on the sky (5.4~pc) and have a 3D separation
of 8.9~pc. 
The corrected proper motion differences (\apx 12 \pmt\ 5
\masyr) may seem a bit too large to accept this system as a real
wide-binary candidate, but this difference is time dependent.  
Capella is an almost equal-mass spectroscopic binary with component 
masses 2.466 \& 2.443 \Msun. 
It has, 12\am away, known companion WDS 05167+4600 HL which
comprises the M1V star GJ~195~A (V=10.16 mag) and the M5 dwarf
GJ~195~B (V=13.7 mag), which are separated by 6\as.  
50~Per itself is paired here with HIP~19255 at a
separation of \apx 15,200 AU (\apx 740\as).  Furthermore, the MSC
identifies both 50~Per and HIP~19255 as possible binaries
themselves, while these systems orbit each other in about one
million years at an equivalent circular orbit speed of \apx0.45
\kms\ (4.8 \masyr). The MSC reports a total mass 3.64 \Msun\ for the
50~Per and HIP~19255 system. The two components of HIP~19255 are
separated by 3\asb87 and orbit each other in 590 years
($v_{orb}\apx$ 4 \kms = 42.5 \masyr). This orbital speed is more
than sufficient to account for the corrected proper motion
difference between 50~Per and HIP~19255. We find the following about
the putative 50~Per binary: 1) The HIP2 and TY2 proper motions for
50~Per differ by 2.7 \pmt\ 1.7 \masyr, 2) HIP finds an acceleration
in the proper motion which \citet{MK2005} estimate to be 5.8 \pmt\ 3
\masyrsq\ (\apx 0.5 \kmsyrsq), and 3) the GCSN reports 8 observations
over a period of 7 years with measurements errors of 0.2 \kms\ and
an ensemble error of 0.6 \kms: this is consistent with an
acceleration of 0.286 \pmt\ 0.06 \kmsyrsq. Thus, both RV and proper
motion data indicate the presence of an unseen companion for 50~Per.
The corrected proper motion and RV differences between
50~Per and Capella are bridged at the observed accelerations of
50~Per after 4 \pmt\ 3 and 17 \pmt\ 4 years, respectively.

The total mass for the Capella/50~Per system is (5.88+3.64)=9.52
\Msun, so that the Jacobi radius is 2.8 pc, or about three times
smaller than the observed separation. Thus, Capella and 50~Per may
be an example of an escaped binary system.

Although we also list the known double, HIP~26779 and HIP~26801, at 509\am 
or 2 pc from Capella as having very high probability for being physically 
related to Capella, unless the radial velocity is just
wrong, we suspect that these are false positives. 
The barycentric velocity  is well established, therefore
the $\sim25$ \kms\ difference in radial velocities is hard to explain unless
this system is just passing by. 

\section{Conclusions}

We have applied a full Bayesian approach to assigning probabilities of
companionship between HIP stars separated by more than 0.01 pc.  
By companionship we mean either bound gravitationally as in a system of small
numbers of stars or co-moving with nearly the same velocity
as in an escaped previously bound component.  
After subtracting the expected numbers of false positives derived
from control experiments,  a population of companions
extending out to 8 pc in separation remains.  
Some of these very wide systems contain
hierarchies of fairly massive stars that extend the tidal
radii out to unusually large distances, but it is likely that others
are recently unbound systems that continue to travel along nearly
the same trajectory. 
While some of these seem to be parts of known nearby moving clusters or 
associations (\eg Tucanae Stream, Hyades Stream, UMa Moving Cluster, 
$\beta$ Pic Moving Group, and TW Hydrae Association), this procedure brings 
to focus even higher density knots within them,  which should be far more
persistant than the rest of the association either as a bound system or a
tight stream.
The amount of time after breakup of an open cluster or binary system for 
which companions stay in close proximity may be an important constraint on 
the mass and distribution of dark matter candidates such as dark subhalos.

Our statistical method finds both many highly significant pairings that 
are missed by previous techniques and assigns reasonable probabilities for
companions even in regions previously considered too complicated or
crowded.  
In  the 1 -- 100 pc distance range, we find altogether $\sim$222 high 
probability HIP-HIP companions with
separations between 0.01 -- 1 pc, and we find strong evidence for a
population of companions separated by 1 -- 8 pc with $\sim314$ stars.  
In just the 0 -- 25 pc range, we find $\sim$34 companions with separations 
0.01 -- 1 pc and $\sim$50 companions with separations 1 pc -- 8 pc.
Our preliminary investigations do not show any obvious trend for the 
excess of wide/escaped
binaries along the Galactic rotation direction.

As displayed in Fig.~\ref{fig:vr}, we find good
agreement between the radial velocities of the primary and the
corrected RV of the candidate companions: 56\% have velocity
differences $<6.8$ \kms\ (about $3\sigma$). For comparison,
the distribution of RV differences of random nearby HIP-HIP
pairs closely resembles a zero-centered Gaussian with a dispersion
of \apx37.5 \kms, which leads to a 15.8\% chance that a random pair
would have a velocity difference as small as 6.8 \kms.  In addition,
unresolved spectroscopic binaries can induce RV differences of
order 10 -- 30 \kms, and so pairs with substantial RV differences
might in fact be physically associated. Therefore, the fraction of pairs
with small true RV differences could be significanly higher.

In some individual cases, such as Capella two of its candidate
companions (HIP~26779 \& 26801), the RV data indicates that the
candidates are not real. Out of 426 candidates with radial velocities, we find 187 (44\%)
pairs for which the corrected radial velocities deviate by more than
three times the errors: most of these systems might qualify as false
positives.  This rate of potential false positives agrees excellently with the rate identified in the control experiments 
(Fig.~\ref{fig:hip_pc} and Table 5) of 428 control candidates out of 964 HIP candidates 
or 44\%.

\begin{figure}
\epsscale{1.0}
\plottwo{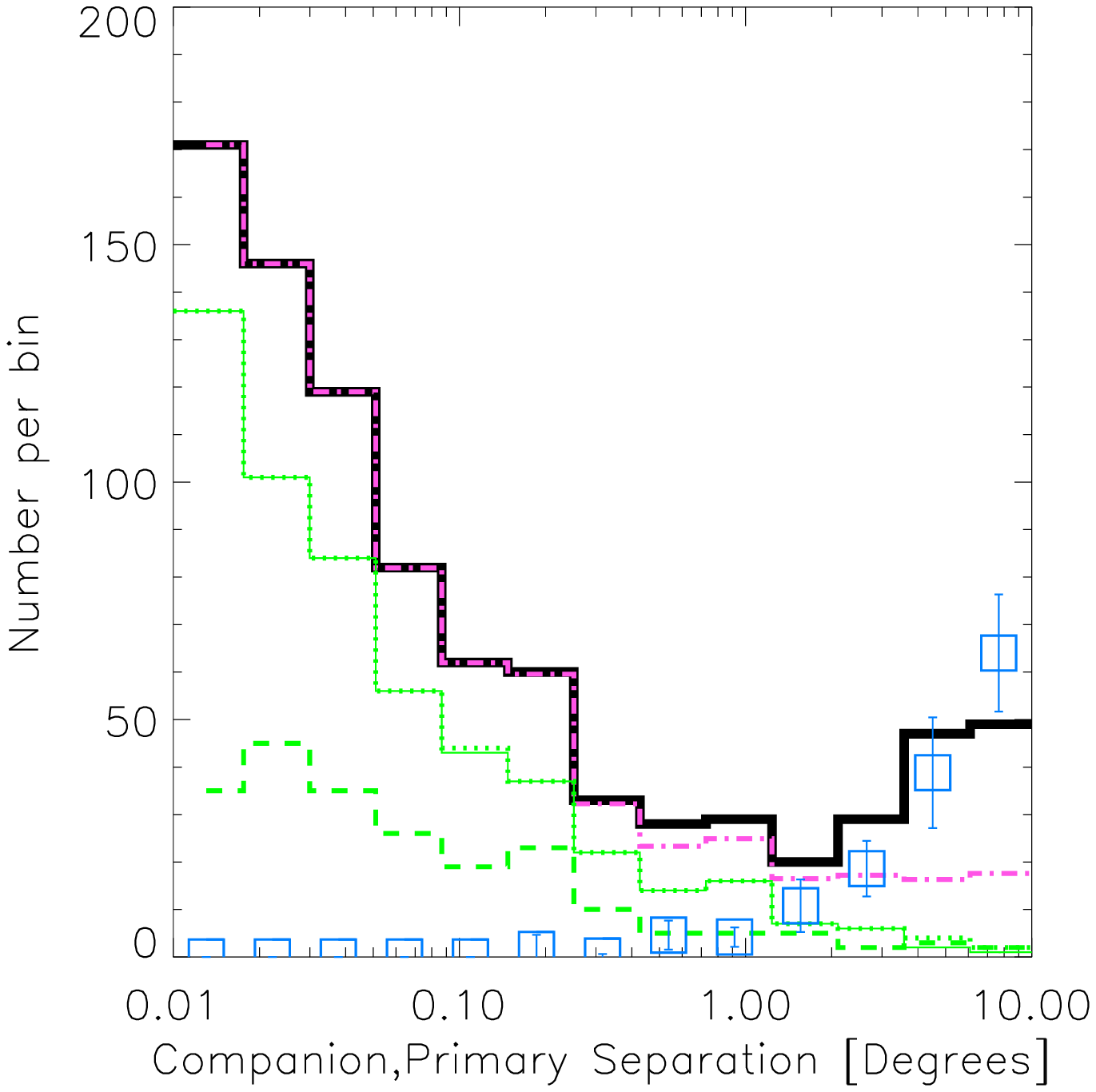}{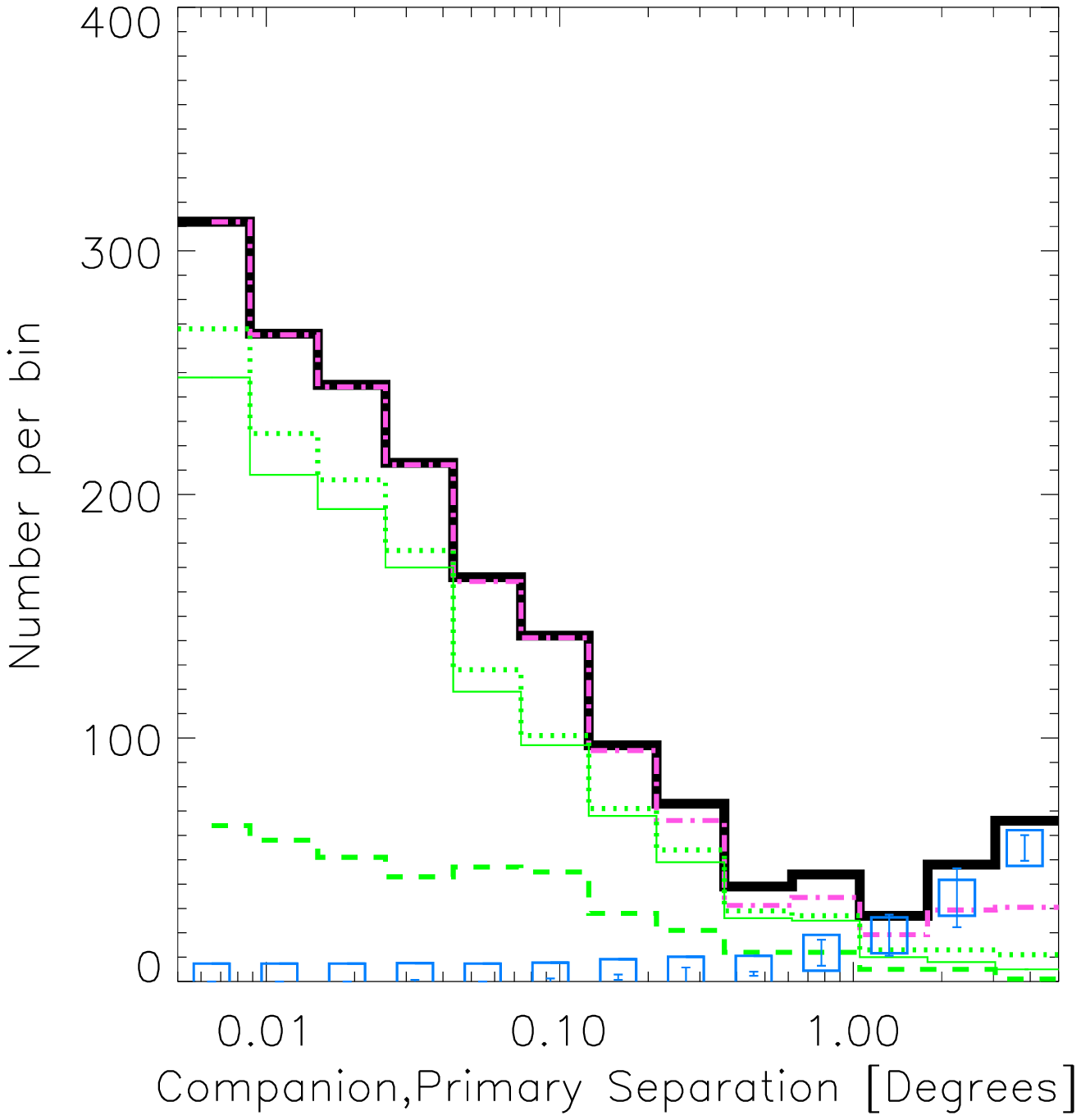}
\caption{
{\it Simulated Number vs. Angular Separation with DM91 Binary Distribution}
 - Results of a simulation with about 50\% of the stars being physically 
related companions that follow the DM91 binary distribution.
The left plot is for primaries between 25 -- 50 pc, and the right is for 50 -- 100 pc.
The green dotted line shows the number of companions per logarithmic
separation bin created in the simulation within the distance interval.  
The black solid line
shows the number of companion candidates found with probabilities $>10$\%.
The green thin solid line gives the number of these that are correct
primary-companion associations and the lower green dashed line gives the
number where the primary was missed but two companions are
scribed to be a primary-companion pair.   
The average over 4 rethrow control experiments plus a reversal 
of Galactic latitude is shown as blue
squares with error bars.  
The purple dashed-dotted line is the sum of the probabilities 
of companions within each bin.
\label{fig:sim_Deg}}
\end{figure}

\begin{figure}
\epsscale{1.0}
\plottwo{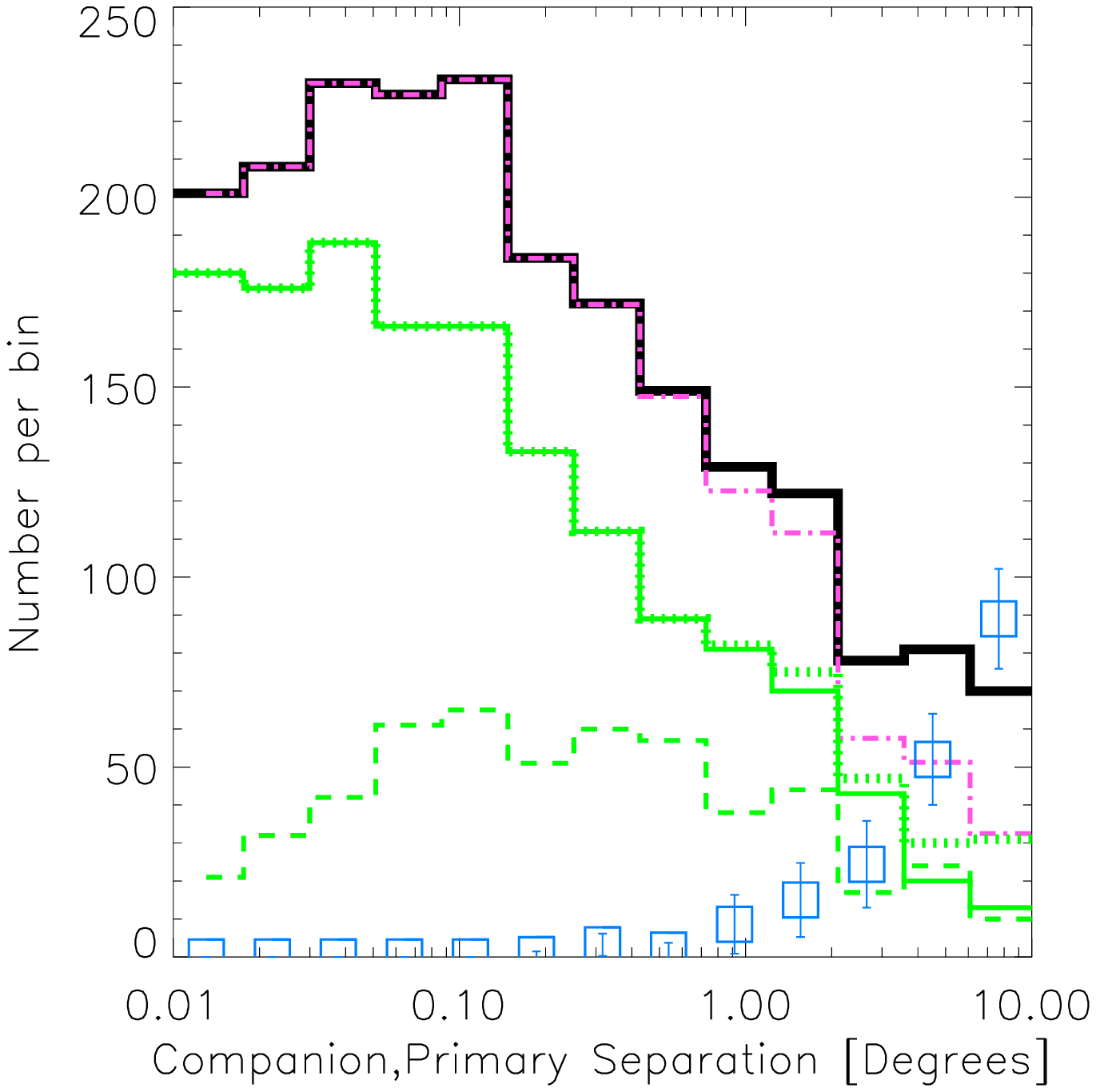}{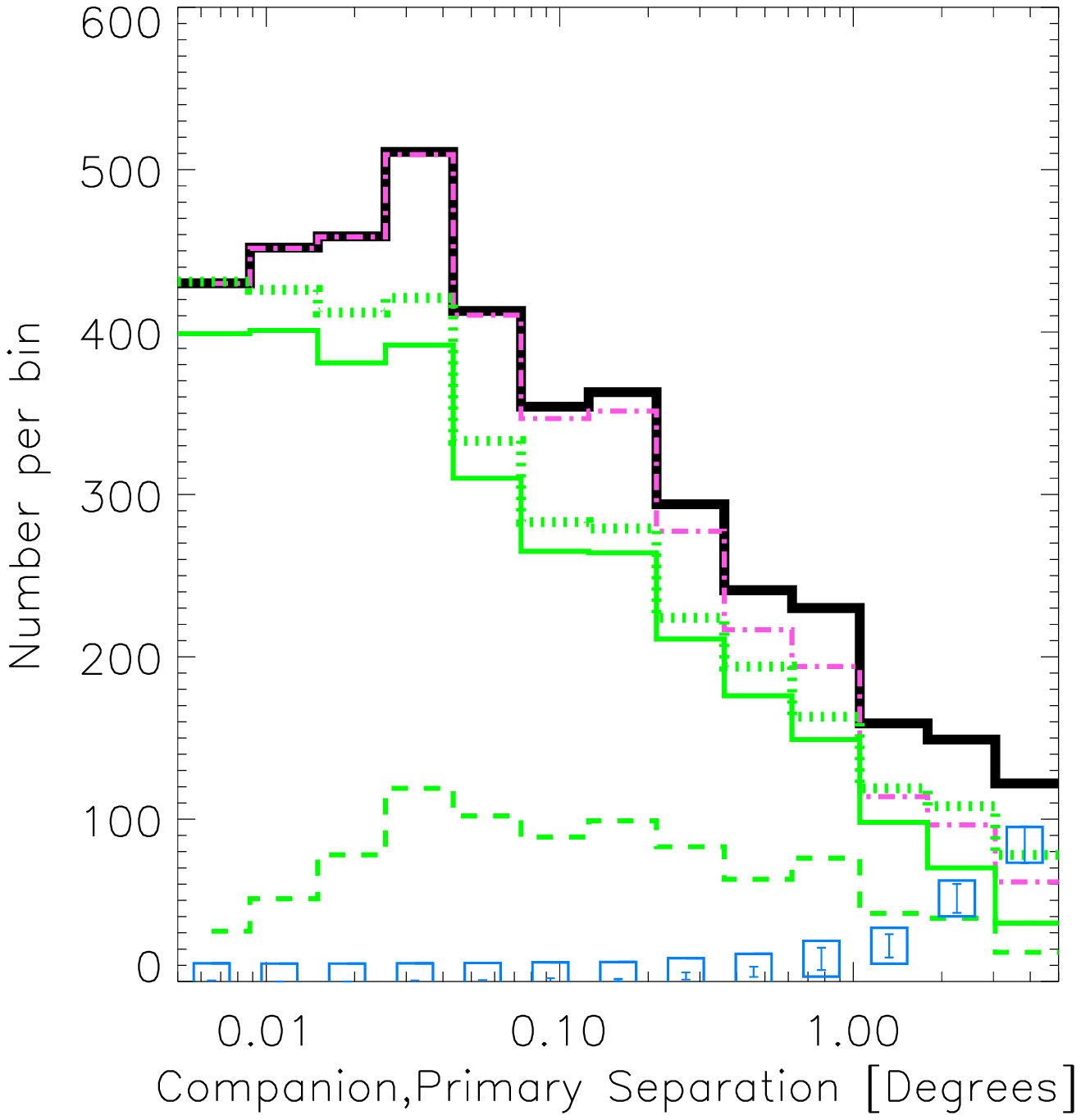}
\caption{
{\it Simulated Number vs. Angular Separation with Expanded DM91 Binary
Distribution} -  Same as previous figure, but periods of the bound systems
were multiplied by 100 to produce more companions at much larger radii.
Now one can compare at very wide separations how many primary-companions are
correctly caught (green thin solid line) compared to 
how many are in the simulation (green dotted line).
\label{fig:sim_Deg_100}}
\end{figure}

\begin{figure}
\epsscale{1.0}
\plottwo{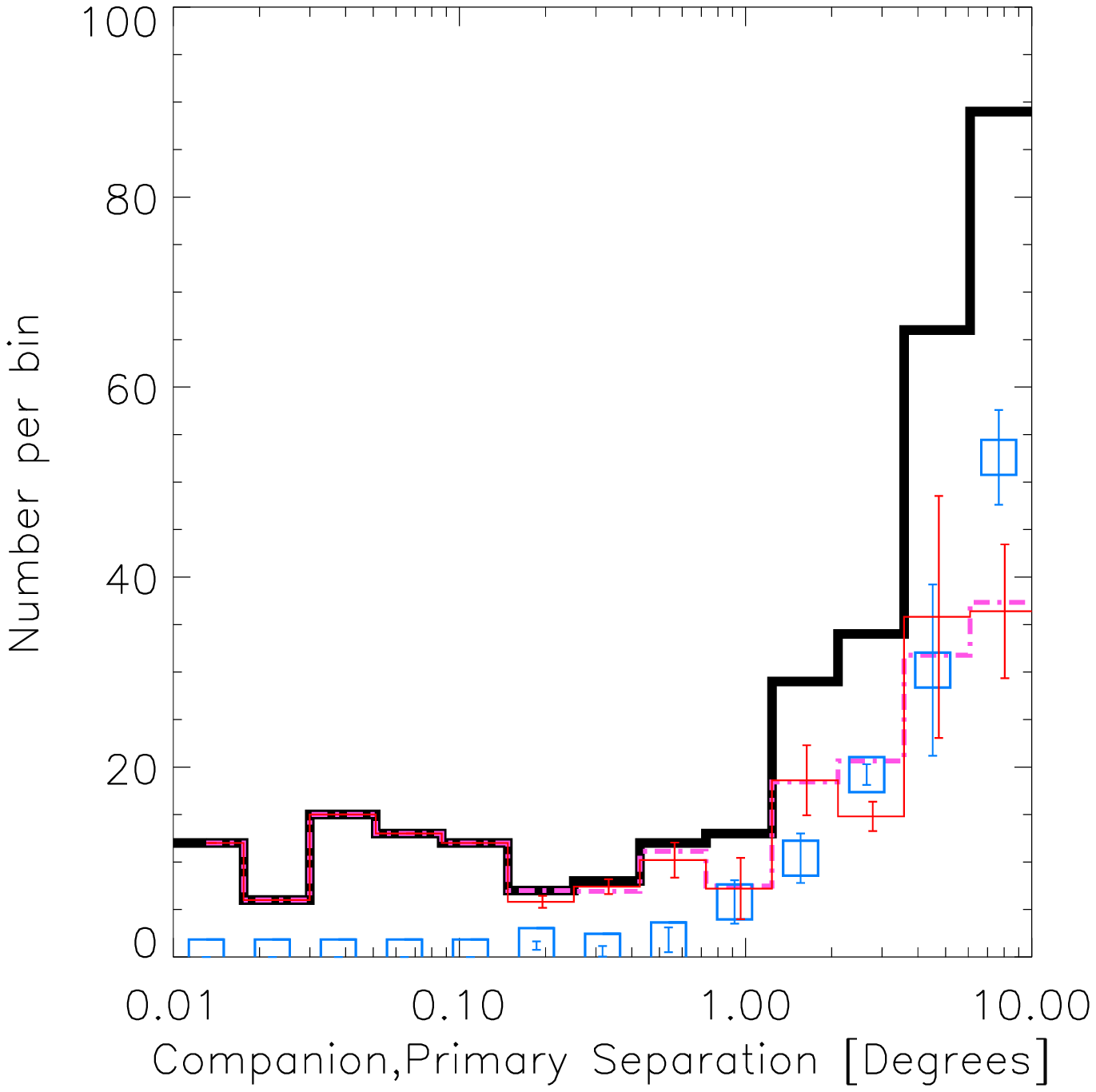}{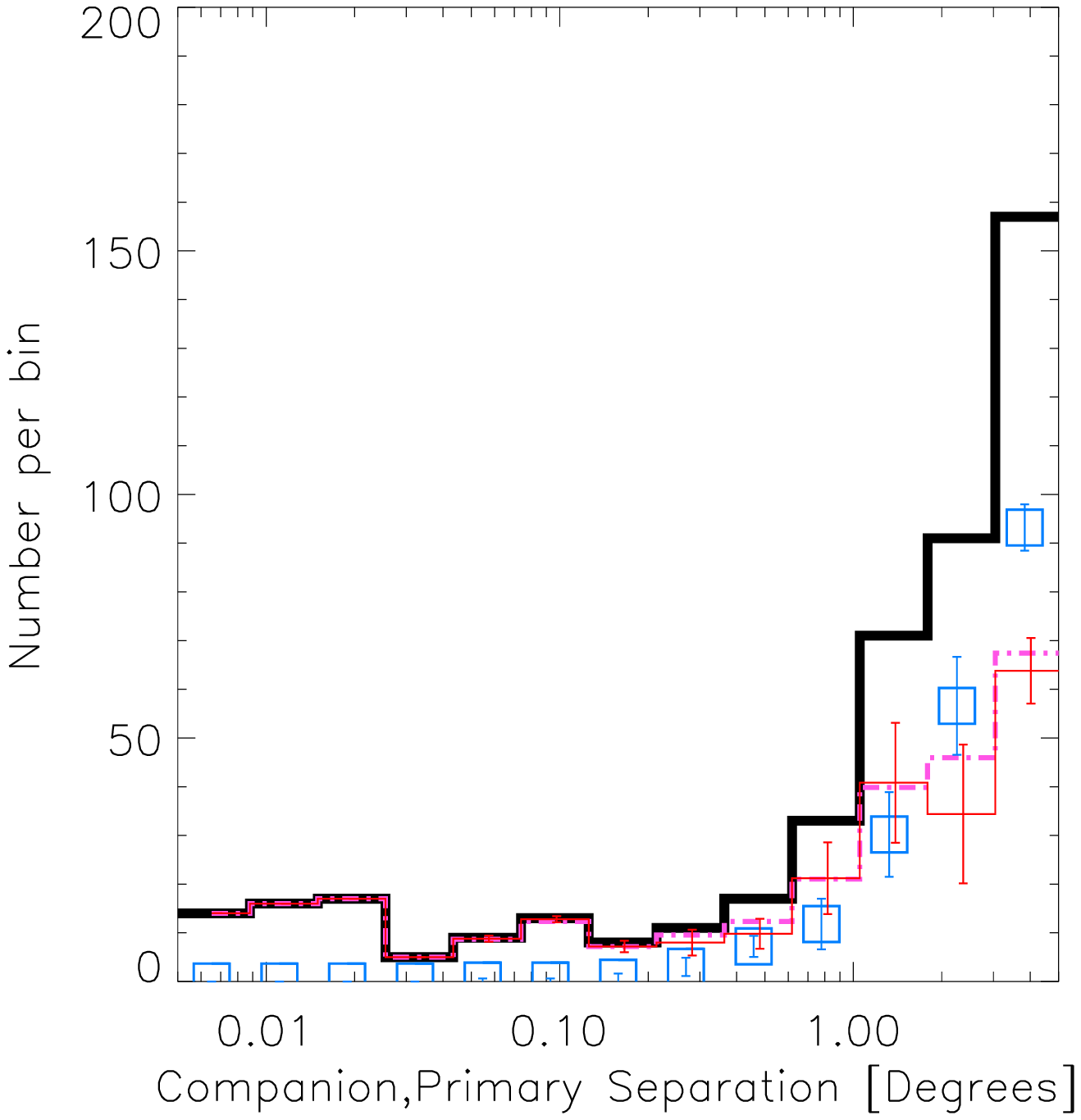}
\caption{
{\it Number vs. Angular Separation of Hipparcos Companions} - results 
of the Hipparcos Catalogue for primaries
between 25 - 50 pc (left) and for 50 - 100 pc (right).  
The black thick solid line shows the number of companions found with 
probabilities $>10$\%.  The averages over 4 rethrow control experiments
and 1 reversal of Galactic latitude are shown as blue squares with error bars.  
The red thin line with error bars shows the observed minus the averages of the control experiments and
therefore provides an estimate of the number of real companions in the
Hipparcos Catalogue.  The purple dash-dotted line is the sum of the
companions' 
probabilities within each bin. Note how different this figure is from the
previous figures of simulations.
\label{fig:hip_Deg}}
\end{figure}

\begin{figure}
\epsscale{1.0}
\plottwo{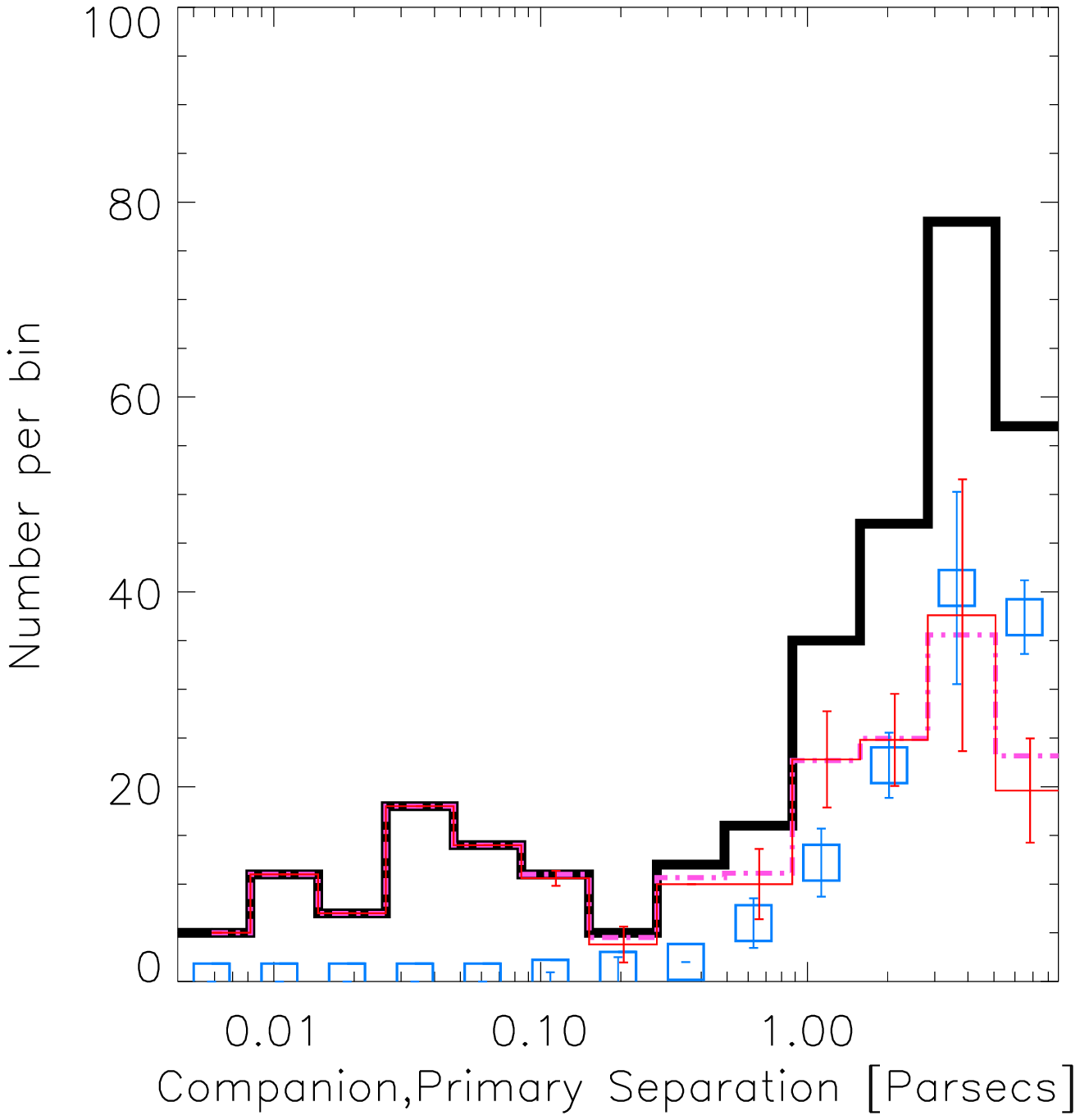}{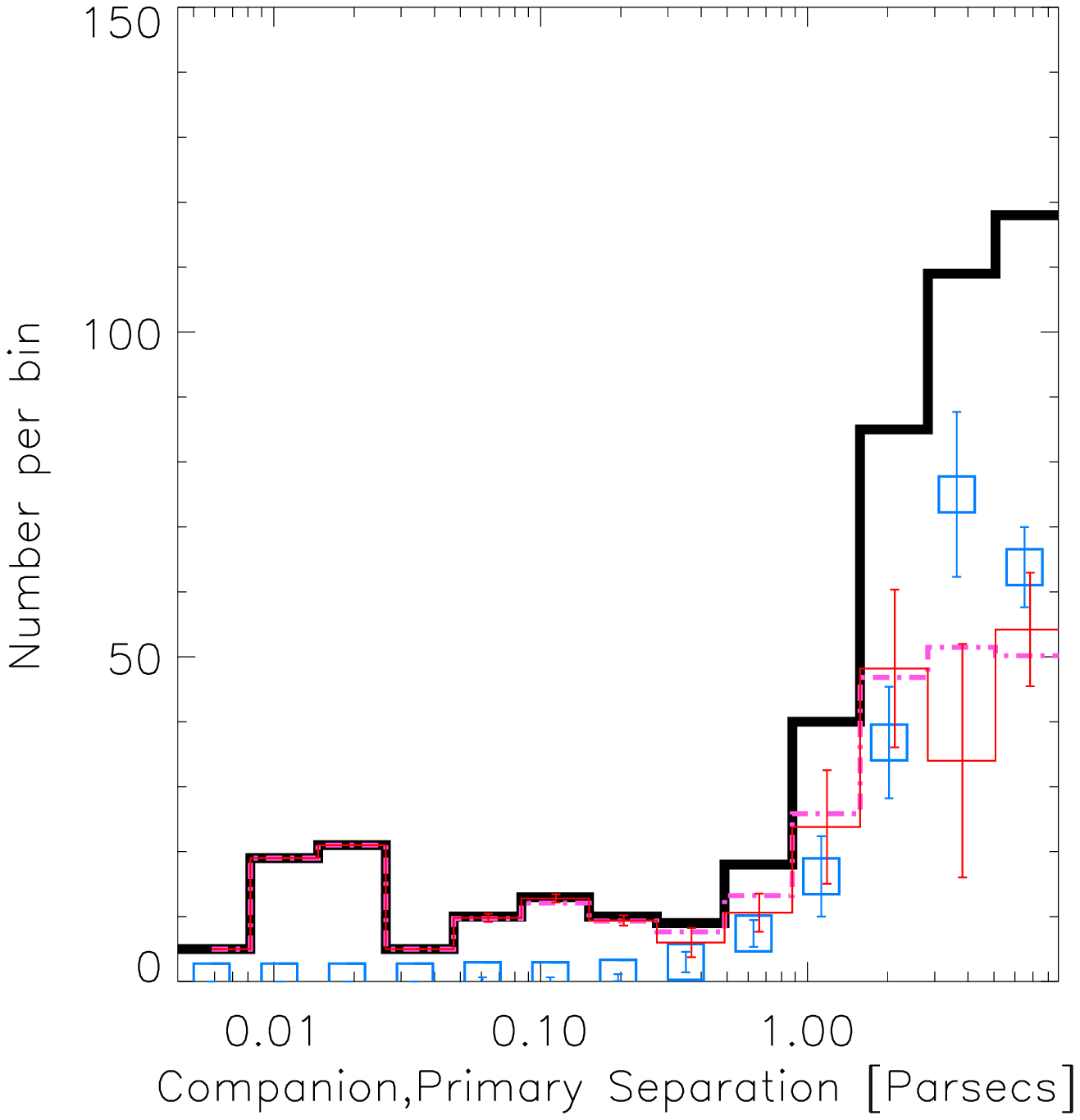}
\caption{
{\it Number vs. Physical Separation of Hipparcos Companions} -  Same as in 
previous figure, but bins are now in logarithmic separation in parsecs
based on parallax measurements of primaries.
\label{fig:hip_pc}}
\end{figure}

\begin{figure}
\epsscale{1.0}
\plottwo{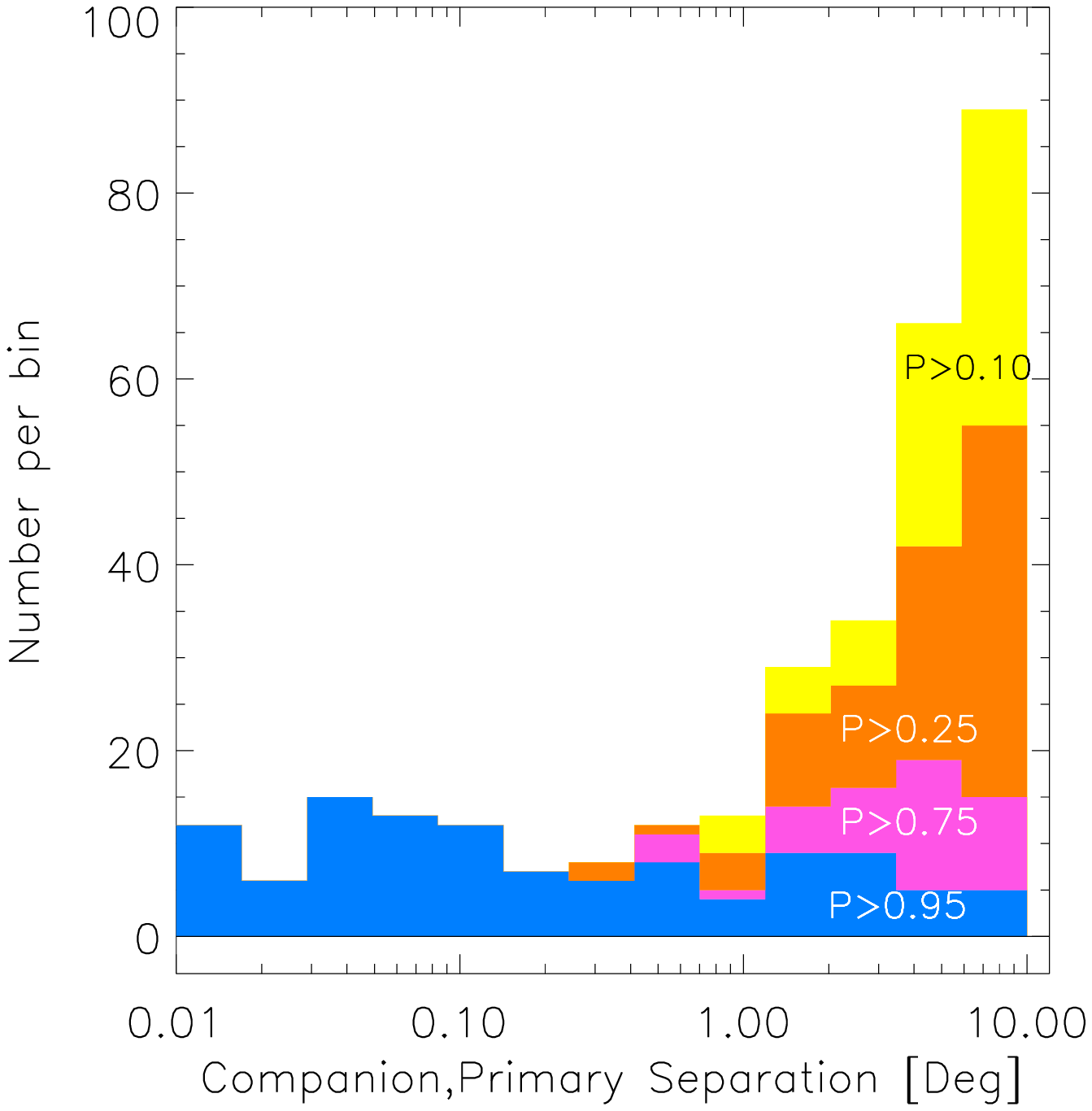}{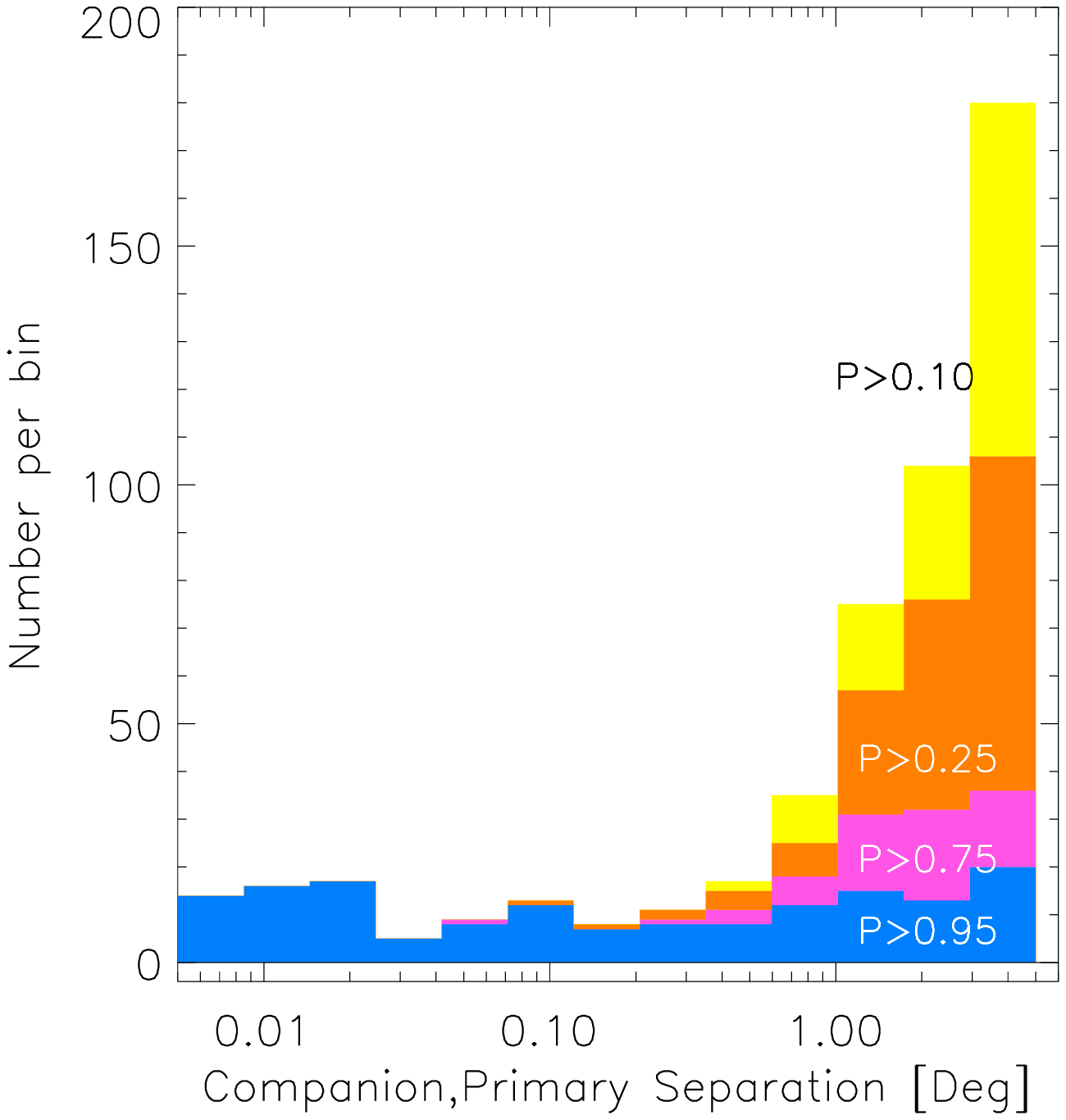}
\caption{
{\it Number and Probability vs. Angular Separation of Hipparcos Companions} - The companion
separation histogram showing the contributions from
different probability ranges.  The contribution from probabilities
$>0.95$ is blue, $0.75 < P < 0.95$ is green, $0.25 < P < 0.75$ is
orange, and $0.1 < P < 0.25$ is yellow.  The left is for primaries
between 25 - 50 pc, and the right is for 50 - 100 pc.
\label{fig:pseps}}
\end{figure}

Figs.~\ref{fig:sim_Deg} and \ref{fig:hip_Deg} indicate
that the classical log-normal period distribution (DM91) results in
a distribution of pair separations that is very different from 
the one we observe. 
To double check this, we use another simulation,
where {\it each} HIP star is assigned a secondary with a mass drawn
from the initial mass function. The absolute magnitudes
for these simulated stars are estimated by applying the inverse of the procedure
outlined in \S\S\ref{sec:Stellar_Masses} above. We then use the
magnitude-completeness function of HIP (as determined by comparing it to the TY2 magnitude counts) to decide to accept or
reject a given simulated secondary. While this procedure
more-or-less reproduces the {\it number} of known HIP-HIP binaries,
the distribution of separations resembles the results obtained from
our simulation (Fig.~\ref{fig:sim_Deg}), but at lower amplitude. 
An attempt was made to match more closely our observed distribution by changing
the location and width of the peak of the log-normal period
distribution, and the form of the IMF. None of these experiments
succeeded. We tentatively conclude that the observed distribution of separations
is incompatible with the log-normal period distribution of nearby
G-type field stars as observed by \citet{DM1991}. 
However, we also must acknowledge that the non-random selection of fainter 
stars in HIP and the fact that this catalog is magnitude limited 
renders extractions of overall statistics on true binarity rates quite uncertain. 

The very wide systems found here are
all smaller than 6.2 $r_J$ and seem to be distributed
as suggested by \citet{Jiang_tremaine_2010}, with a
minimum at about $r_J$, and a rising population of escaped
companions at larger separations.
  The relative velocities are a more
  stringent criterion: from Fig.~6 of \citet{Jiang_tremaine_2010}, we
  infer that escaped binaries should have $\Delta v/v_J \la
  30$.  About 72\% of the our systems satisfy this criterion.
  Including the observational errors, 89\% (98\%) of our very wide
  systems satisfy the criterion within 1$\sigma$
  (2$\sigma$). Thus, we are confident that most of these systems qualify as 
  bona fide escaped bound systems.

  However, they may not have begun as simple binary sytems.
  Other possible sources are the remnants of
  dissolving low-density clusters of stars.
  \citet{Kouw_WideBinForm_2010} show that dissolving clusters can
  produce very wide binaries whose separation can easily reach parsecs,
  and even have rising distributions at separations around one parsec. In
  fact, \citet{Kouw_WideBinForm_2010} argue that the size of the
  semi-major axis of {\it young} wide binaries is similar to the
  initial size of the cluster from which they formed.  Our systems are
  moderately young (typical mass 1.5\Msun), and so the bound ones may still
  reflect the size of their birth places. Another prediction
  \citet{Kouw_WideBinForm_2010} make is that very wide binaries should
  be preferentially hierarchical with each of the wide components being
  binaries by themselves. Indeed there are observations that are consistent
  with this prediction \citep{MZH_2008, Caballero_2009,
    Caballero_2010}. 
    For the few systems that we have thus far tried to 
  collect possible companions from the literature, we do indeed find a
  preponderance of hierarchical systems.

We have discovered some hitherto unnoticed pairing of very
nearby stars  and a large number of pairings at
record separations.
Subhalos would greatly accelerate the disruption of wide binaries if they are 
an important contributor to the small scale potential locally in
the Galaxy.
Unfortunately, the distribution of subhalos in the Galaxy at the solar radius is not yet well
predicted by N-body or hydrodynamical simulations \citep{Gan_etal_2010}.
Since companionship of escaped binary companions at large separations is very
fragile, requiring comoving
velocities to remain $\lesssim$ 1  \kms, statistics 
on the number of very wide companions and their ages should lead to useful limits on the masses
and number densities of dark matter subhalos. 

Statistical algorithms for ascertaining probabilities of association
and/or boundedness in large astrometric surveys with high precision
will become more effective as larger and more precise astrometric
surveys come along, such as Pan-Starrs \citep{PSTARRS_Astrometry}, LSST
\citep{LSST_Astrometry}, and Gaia \citep{Gaia_2002,Gaia_2008}. 
The astrometric data from Gaia will be about
10 times better than the positional data obtained from the former
two ground-based projects. 
When available, such data will enable a full mapping of the six-dimensional
phase-space distribution of any potential physical binary. 
To facilitate the analysis of these future catalogs, as
well as the analysis of the existing astrometric catalogs, we
investigated the reliability of Bayesian algorithms for providing
realistic probabilities of extremely wide companions and found it to be quite 
successful even when implemented in only simplified form.   
In the future we hope to make more complete statistical use of measurement 
errors, magnitude binning, and incorporation of radial velocities 
and to apply these to the full TY2.
 
\vspace*{1em}

{\it Acknowledgements:} We thank Scott Tremaine for some useful discussions and the referee for useful improvements.  EJS and RPO were partially supported by the SIM 
project which was carried out for NASA by JPL under a contract with the 
California Institute of Technology.  In this work we made
  extensive use of the Hipparcos and Tycho-2 Catalogues
  \citep{ESA1997, HIP2}.  This work would have been virtually
  impossible without the aid of the SIMBAD database, operated at the
  ``Centre de Donn{\' e}es astronomiques de Strasbourg'' (CDS). RPO
  was supported by an inheritance from his late father, Dr. Ch. C. J.
  Olling.

\clearpage

\begin{figure}
\epsscale{1.0}
\plottwo{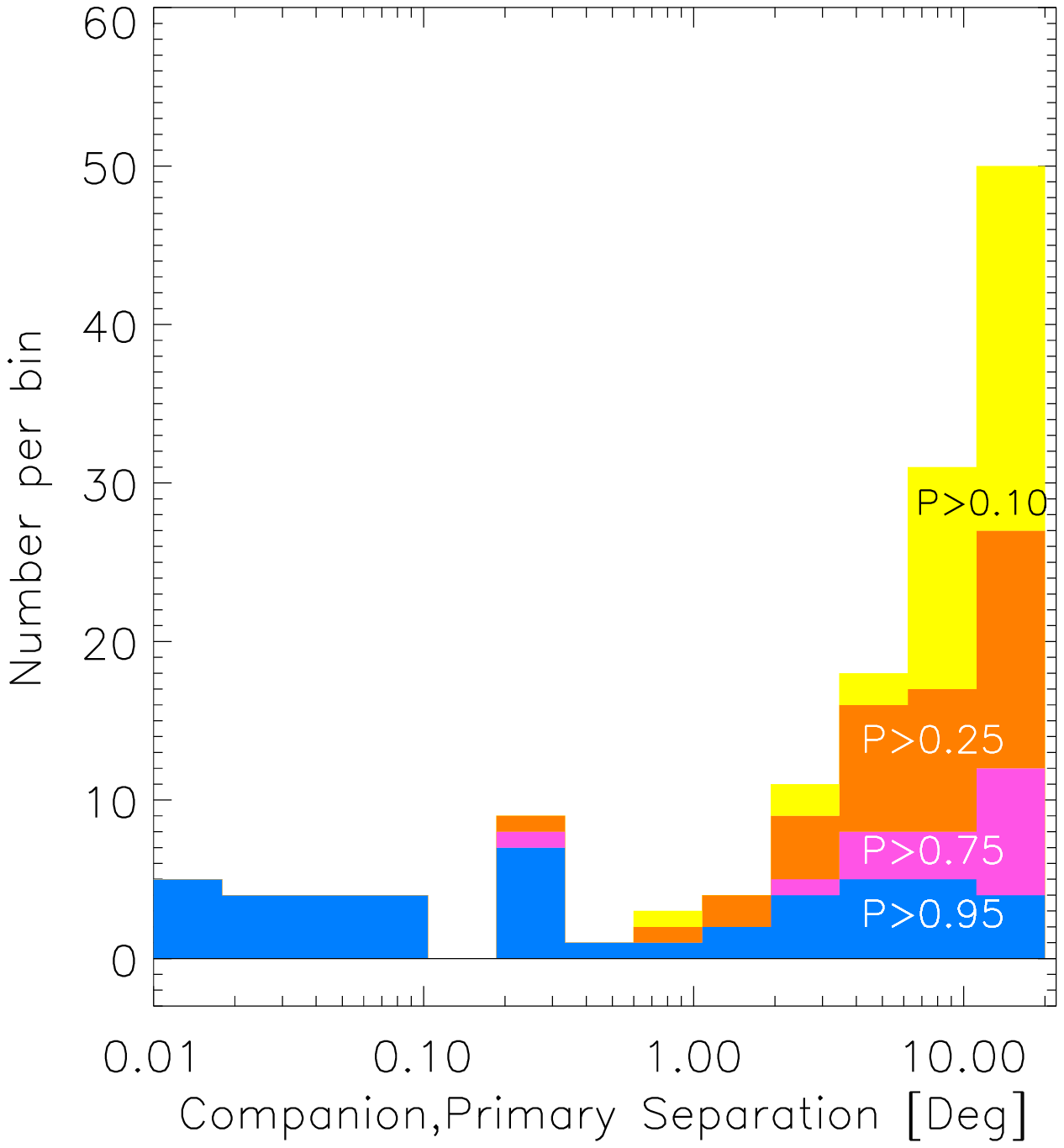}{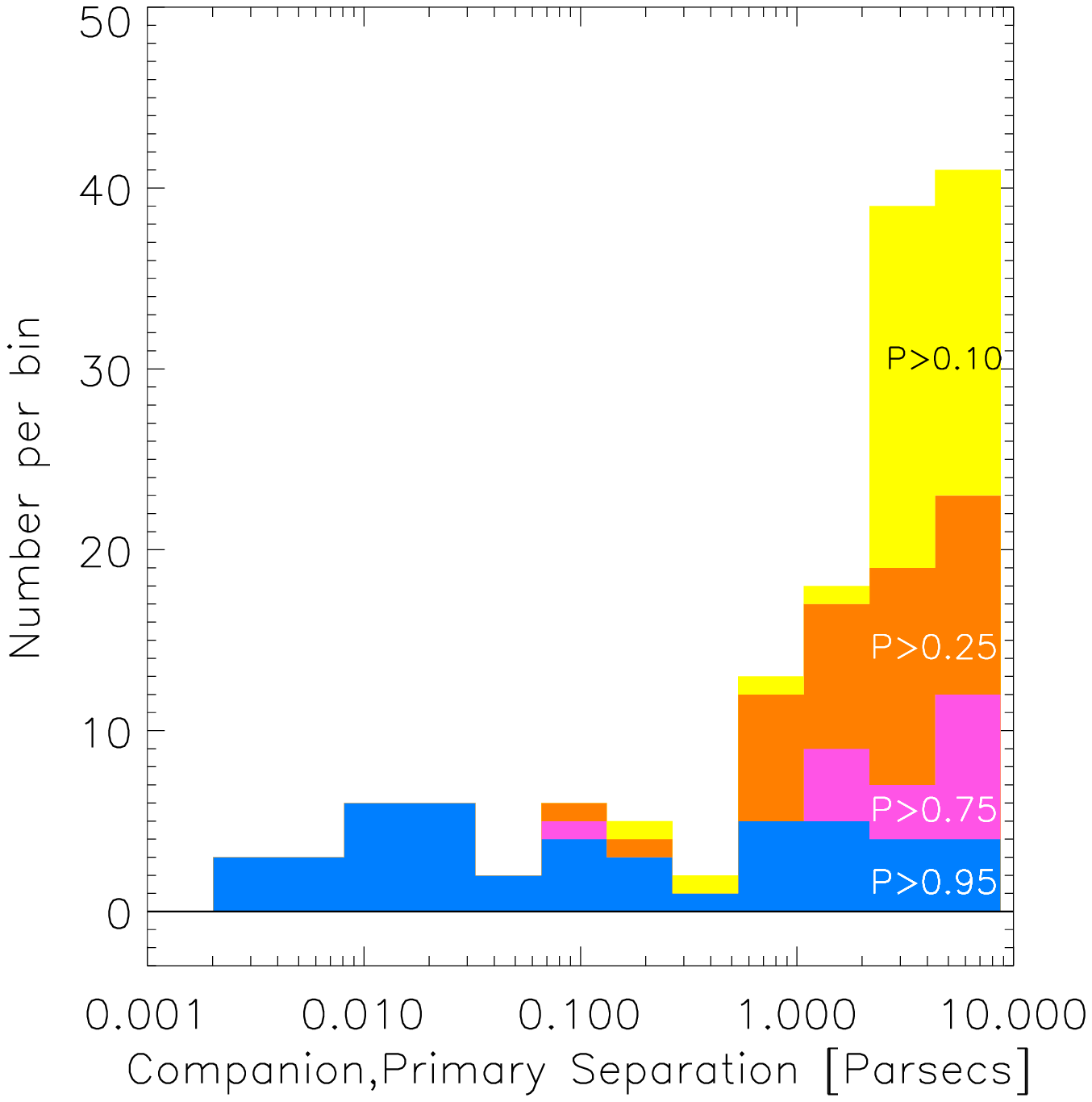}
\caption{
{\it Number and Probability vs. Separation, 0 -- 25 pc distances} - 
The companion
separation histogram for primaries within 25 pc showing the contributions from
different probability ranges as in previous figure.  
The left plot shows angular separation and the right is angular separation 
times distance to give projected separation in parsecs. 
\label{fig:pseps0-25}}
\end{figure}

\begin{figure}
\epsscale{0.9}
\plotone{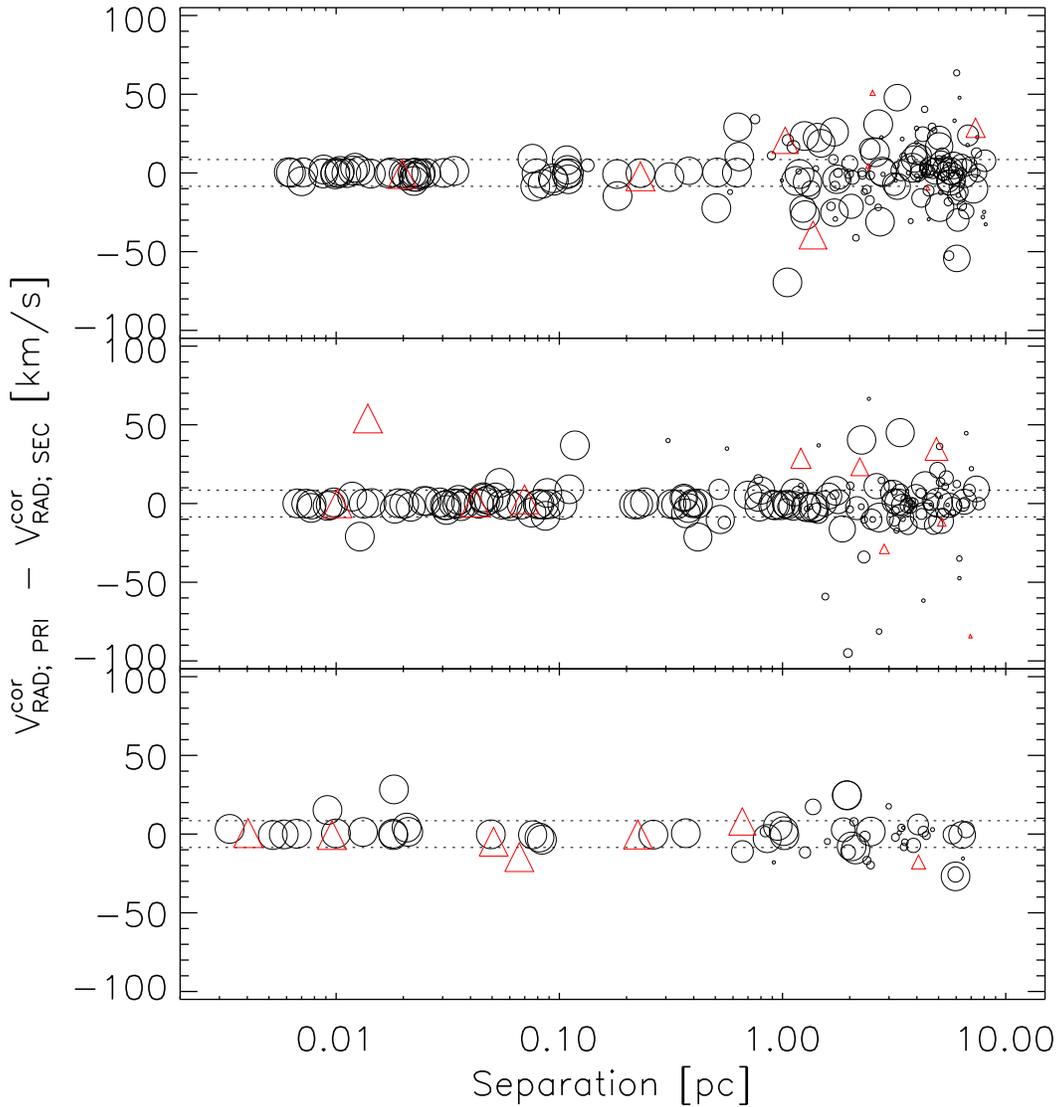}
\caption{
{\it Radial Velocity Differences} - Corrected radial velocity differences vs.
separation in the plane of the sky in parsecs for Hipparcos stars in which
radial velocities for both primary and companion are available. Triangles are
used for stars listed in the SIMBAD database to be some kind of binary.  
The size of the symbol is proportional to the probability of companionship 
down to P = 0.1. 
The primary's space motion is transformed to the direction of the primary
to obtain the expected radial velocity of each companion and then the
companions' radial velocities are subtracted.   
The dispersion is much less than expected if randomly drawn from the 
radial velocities distribution in the Solar Neighborhood.  
Since the difference in radial velocity 
was not used in calculating probabilities, the good agreement of the
majority of systems provides high confidence that
the technique can find physical companions. 
The top frame is for primaries between 50 -- 100 pc, middle
is 25 -- 50 pc, and the bottom is 0 -- 25 pc.
\label{fig:vr}}
\end{figure}

\begin{figure}
\epsscale{0.9}
\plotone{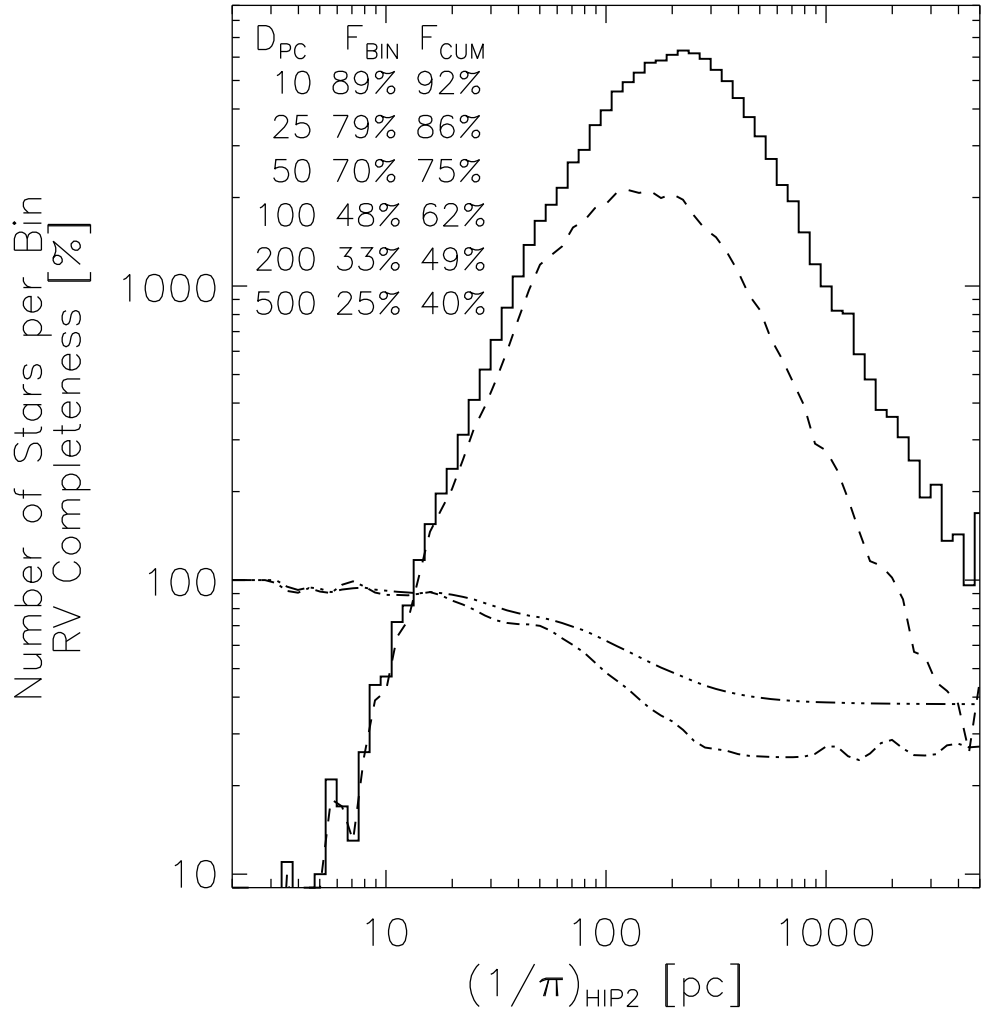}

\caption{{\it RV statistics in of Hipparcos Stars} - 
  The number of HIP stars (per bin)
  as a function of the inverse of $\pi$ as listed in HIP2
  (histogram; positive parallaxes only).  The thick dashed line is the
  number of stars in our radial velocity database. We show two
  versions of the completeness of our radial velocity catalog: 1) the
  completeness per bin ($F_{BIN}$; thick dash-dotted line) is the
  ratio of the number of stars with RVs to the number of
  stars, and 2) the cumulative completeness ($F_{cum}$;
  dash-triple-dotted line) which is the total number of stars with RVs
  up to distance $1/\pi$, divided by the total number of stars out to
  the same distance. We list both completeness values at some
  representative distances. Note that, since the typical parallax
  errors are of order 1 -- 2 \mas, distances beyond a few hundred
  parsec are not very well determined.
\label{fig:RV_completeness}}
\end{figure}

\begin{figure}
\epsscale{1.0}
\plotone{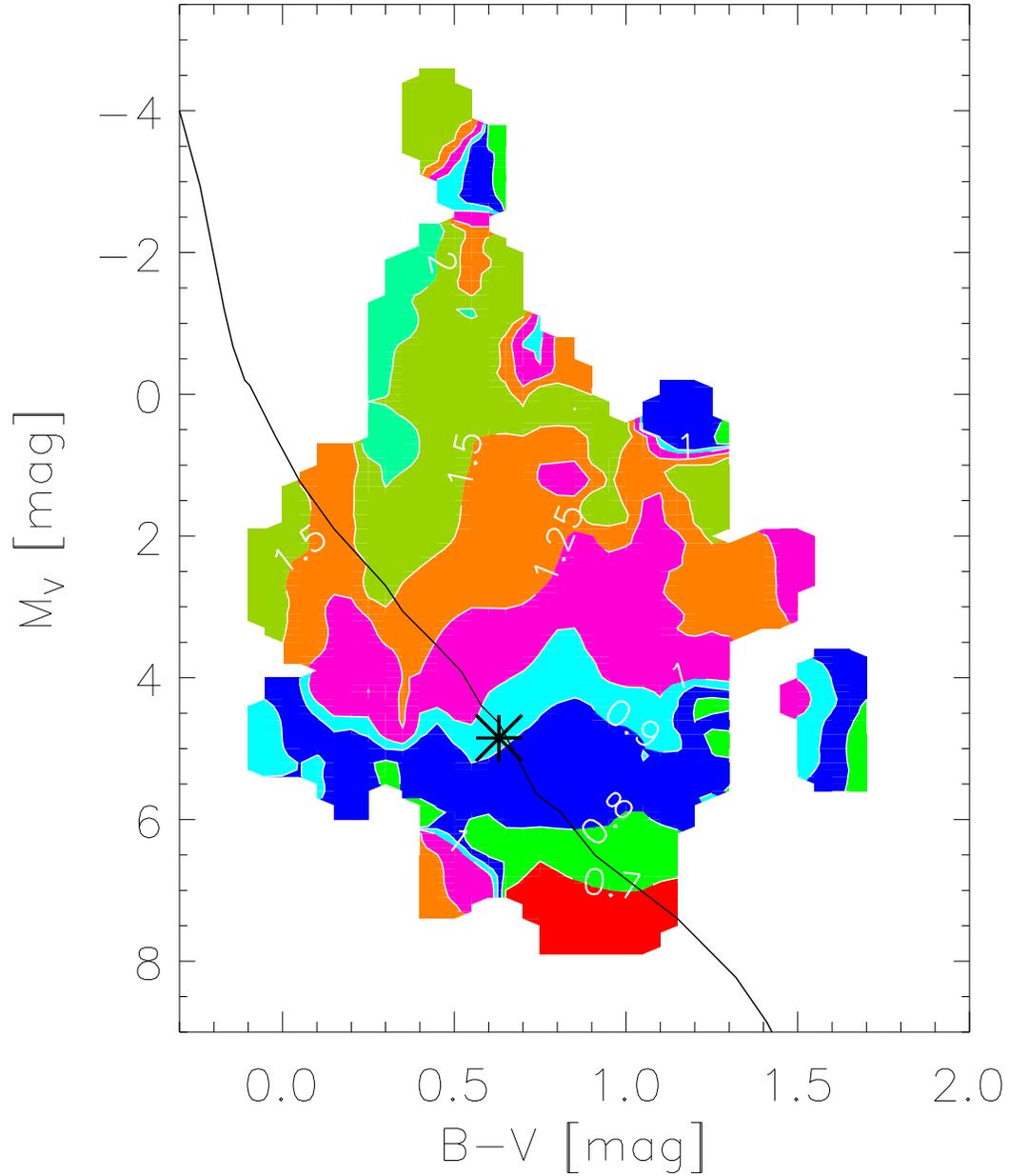}
\caption{{\it GCSN Mass Map} - The mass as determined from the GCSN
  as a function of (B-V)-color (abscissa) and absolute magnitude
  (ordinate). The contours are labeled by the mass values The solid
  line represents the main sequence, and the star-symbol the location
  of the Sun. The fact that the Sun does not lie in the ${\cal M}$=1
  contour, but at ${\cal M}=0.925$, indicates that the errors on the
  inferred masses are not negligible: about 7.5\% on the
  MS, about 15\% below the MS and up to 20\% towards the top and
  right-hand side of the map.}  
\label{fig:Stellar_Masses}
\end{figure}
\clearpage

\begin{deluxetable}{cccccccccccccc}
\tabletypesize{\small}
\rotate
\tablecaption{Parameters for Probability Estimator \label{tbl:params}}

\tablewidth{0pt}
\tablehead{
\colhead{Distance}&
\colhead{$\Delta\mu_{lim}$} & 
\colhead{$\Delta\mu_{inner}$}& 
\colhead{$\Delta\mu_{outer}$} & 
\colhead{$\Delta d_{max}$} & 
\colhead{$\theta_{inner}$}&
\colhead{$\theta_{outer}$}& 
\colhead{$\mu_0$}& 
\colhead{$\alpha_0$} & 
\colhead{$\Delta \mu_1$}& 
\colhead{$\alpha_1$} & 
\colhead{$N_f^{max}$}\\

\colhead{pc} & 
\colhead{\masyr} & 
\colhead{\masyr} & 
\colhead{\masyr} & 
\colhead{pc} & 
\colhead{deg} &
\colhead{deg}& 
\colhead{\masyr} & 
\colhead{$\cdots$} & 
\colhead{\masyr}  & 
\colhead{$\cdots$} & 
\colhead{$\cdots$}
}
\startdata
0 -- 25     & 35 	& 45 & 100 	 & 20 & 0.01  	&   20 	&  10.0  & 0.97	    &   7.13	& 0.97  & None \\
25 -- 50    & 25 	& 35 & 75  	& 15  &0.01  	&   10 	&  3.44  & 0.93	    &	3.39	& 0.78	&12	\\
50 -- 100   & 10  	& 20 & 50  	& 20  &0.005  	&   5 	&  2.09  & 1.06	    &	2.28	& 1.15	&10	\\
\enddata

\end{deluxetable}

\begin{deluxetable}{rrrrrrrrccc}
\tabletypesize{\small}
\rotate
\tablecaption{Attrition of Primary Stars
\label{tbl:attrition}}
\tablewidth{0pt}
\tablehead{
                  
&                 
& 
\colhead{Has}&
\colhead{$N_f$}& 
\colhead{Has}&
\colhead{\# of Cs}&
\colhead{\# of Ps}& 
\colhead{$N_c$ -}&
\colhead{($N_c -$} &                  
& \\ 
\colhead{Distance} &
\colhead{$V<10$} & 
\colhead{Field} & 
\colhead{$< N_{f,max}$} &
\colhead{Cs} &
\colhead{$P>0.1$}   &
\colhead{$P_p>0.1$} &
\colhead{$N_c^{ctrl}$} &
\colhead{$N_c^{ctrl})(<1pc)$} &
\colhead{$\sum P$} & 
\colhead{$\sum P_{<1pc}$}
}
\startdata
0 -- 25           & 1041            & 786                 & 786                     &  369              & 144                 &  111          &  84                    & 34 &  85.7            &  26.9 \\\hline
25 -- 50   		  & 4152            & 2916                & 2889
&  1119              & 316                 & 265           &   194     &  89             & 198.7            &  72.5	\\\hline
50 -- 100         & 14064           & 9018                & 8861                    & 2090              & 504                 & 464           &   258        &  99          & 301.3            & 106.5 \\
\hline
\hline
Total & 19257 & 12720 & 12536 & 3578 & 964 & 840 & 536 & 222 & 585.7 & 205.9\\
\enddata
\tablecomments{Column 1 - distance interval of the row, 
Column 2 - Number of stars brighter than V=10, 
Column 3 - Number left after removing stars with no field stars or candidate companions, 
Column 4 - Number left after removing those with too high field phase-space density, 
Column 5 - Number left after removing those with no candidates,
Column 6 -  Same as 5 but for separation < 1 pc only  
Column 7 - Number of candidates with $P>0.1$, 
Column 8 - Number of primaries with at least 1 candidate with $P>0.1$, 
Column 9 - Number of candidates  - Number of candidates in control experiments with $P>0.1$, 
Column 10 - $\sum P$ for candidates with $P > 0.1$, 
Column 11 - $\sum P$ for candidates with $P > 0.1$ and separated by less than 1 pc.} 

\end{deluxetable}

\input{Table_Binaries_0025.tex}

\clearpage
\input{Table_Binaries_2550.tex}

\clearpage
\input{Table_Binaries_5099.tex}

\end{document}

%% file: Table_Binaries_0025.tex
\textwidth=18.00cm
\begin{deluxetable}{rrrrrlrrrrrrrrrrrl}
\tabletypesize{\scriptsize}
\rotate
\tablecaption{\label{tbl:25pc}  Candidate Companions for 111 Candidate Primaries within  25 pc.}
\tablewidth{0pt}
\tablehead{
\colhead{Indx} &
\colhead{Pri}          & \colhead{Cmp}   & \colhead{$\ell$}            & \colhead{b} & \colhead{V} &\colhead{Type} &
\colhead{$\mu_l$}        & \colhead{$\mu_b$} & \colhead{$\Delta \mu_\ell^{cor}$} &
\colhead{$\Delta \mu_b^{cor}$} & \colhead{d}       &
\colhead{$v_r$}          & \colhead{$\Delta v_r^{cor}$} & \colhead{$\cal M$} &
\colhead{$\Delta\theta d_p$        }&
\colhead{Prob}& \colhead{Comment}\\
 & &  & \colhead{deg} & \colhead{deg} & \colhead{mag} & &
\multicolumn{4}{c}{ \dotfill \masyr \dotfill } &
\colhead{pc} &
\multicolumn{2}{c}{\dotfill \kms \dotfill } &\colhead{\Msun} &
\colhead{pc}& &\\
\colhead{( 1)} & \colhead{( 2)} & \colhead{( 3)} & \colhead{( 4)} & \colhead{( 5)} & \colhead{( 6)} & \colhead{( 7)} &
\colhead{( 8)} & \colhead{( 9)} & \colhead{(10)} & 
\colhead{(11)} & \colhead{(12)} & \colhead{(13)} & \colhead{(14)} & \colhead{(15)} & \colhead{(16)} & \colhead{(17)}
& \colhead{(18)}}
\startdata
  1&\hspace*{-1.5em}   473& \hspace*{-2.0em}      & 114.6&\hspace*{-3.0em} -16.3& \hspace*{-0.5em}  8.20 ( 78)&\hspace*{-0.5em}  K6V:;SpB&\hspace*{-1.0em}  844.4& \hspace*{-1.0em} -312.7& \hspace*{-1.0em}                   &\hspace*{-1.0em}                   &\hspace*{-1.0em}11.3\pmt  0.2&\hspace*{-2.0em}   0.6 & \hspace*{-1.0em}                &\hspace*{-3.0em} 0.51& \hspace*{-1.0em}       & \hspace*{-1.0em}& \hspace*{-1.0em}\\
   &\hspace*{-1.5em}   473& \hspace*{-2.0em}   428& 114.6&\hspace*{-3.0em} -16.3& \hspace*{-0.5em}  9.95      &\hspace*{-0.5em}   M2e;OtH&\hspace*{-1.0em}  826.6& \hspace*{-1.0em} -309.8& \hspace*{-1.0em}   21.8 \pmt  20.3 &\hspace*{-1.0em}   -4.9 \pmt   7.7 &\hspace*{-1.0em}11.3\pmt  0.2&\hspace*{-2.0em}   0.3 & \hspace*{-1.0em}   0.2\pmt   0.6&\hspace*{-3.0em} 0.45& \hspace*{-1.0em} 0.0180& \hspace*{-1.0em}1.000& \hspace*{-1.0em}\\ \hline
  2&\hspace*{-1.5em}   518& \hspace*{-2.0em}      & 117.0&\hspace*{-3.0em}  -3.9& \hspace*{-0.5em}  5.98 ( 41)&\hspace*{-0.5em}   G5V;SpB&\hspace*{-1.0em}  271.7& \hspace*{-1.0em}  -17.6& \hspace*{-1.0em}                   &\hspace*{-1.0em}                   &\hspace*{-1.0em}21.5\pmt  0.3&\hspace*{-2.0em}  -8.0 & \hspace*{-1.0em}                &\hspace*{-3.0em} 0.98& \hspace*{-1.0em}       & \hspace*{-1.0em}& \hspace*{-1.0em}\\
   &\hspace*{-1.5em}   518& \hspace*{-2.0em}   916& 117.7&\hspace*{-3.0em}  -4.1& \hspace*{-0.5em}  9.50      &\hspace*{-0.5em}    K7;UnK&\hspace*{-1.0em}  234.8& \hspace*{-1.0em}  -16.9& \hspace*{-1.0em}  -26.4 \pmt  12.1 &\hspace*{-1.0em}    3.3 \pmt   3.0 &\hspace*{-1.0em}28.1\pmt  1.5&\hspace*{-2.0em}       & \hspace*{-1.0em}                &\hspace*{-3.0em} 0.63& \hspace*{-1.0em} 0.2520& \hspace*{-1.0em}0.690& \hspace*{-1.0em}\\
\enddata
\end{deluxetable}